\documentclass[aps,10pt,prb,longbibliography,superscriptaddress,twocolumn,showkeys,amsmath,amssymb,floatfix]{revtex4-2}

\usepackage[english]{babel}
\usepackage[T1]{fontenc}
\usepackage{amsmath,amssymb,bbm,mathrsfs,bm,braket,color,graphicx,comment,amsfonts,dsfont}
\usepackage{siunitx,physics}
\usepackage[colorlinks,citecolor=blue,urlcolor=blue]{hyperref}
\usepackage[version=4]{mhchem}
\usepackage[normalem]{ulem}

\begin{document}

\title{Stability of Weyl node merging processes under symmetry constraints}

\author{Gabriele Naselli}
\thanks{These two authors contributed equally}
\affiliation{Institute for Theoretical Solid State Physics, IFW Dresden and W\"{u}rzburg-Dresden Cluster of Excellence ct.qmat, Helmholtzstr. 20, 01069 Dresden, Germany}

\author{György Frank}
\thanks{These two authors contributed equally}
\affiliation{Department of Theoretical Physics, Institute of Physics, Budapest University of Technology and Economics, Műegyetem rkp. 3., H-1111 Budapest, Hungary}

\author{Dániel Varjas}
\affiliation{Institute for Theoretical Solid State Physics, IFW Dresden and W\"{u}rzburg-Dresden Cluster of Excellence ct.qmat, Helmholtzstr. 20, 01069 Dresden, Germany}
\affiliation{Department of Theoretical Physics, Institute of Physics, Budapest University of Technology and Economics, Műegyetem rkp. 3., H-1111 Budapest, Hungary}
\affiliation{Max Planck Institute for the Physics of Complex Systems, N\"{o}thnitzer Strasse 38, 01187 Dresden, Germany}

\author{Ion Cosma Fulga}
\affiliation{Institute for Theoretical Solid State Physics, IFW Dresden and W\"{u}rzburg-Dresden Cluster of Excellence ct.qmat, Helmholtzstr. 20, 01069 Dresden, Germany}

\author{Gergő Pintér}
\affiliation{Department of Theoretical Physics, Institute of Physics, Budapest University of Technology and Economics, Műegyetem rkp. 3., H-1111 Budapest, Hungary}

\author{András Pályi}
\affiliation{Department of Theoretical Physics, Institute of Physics, Budapest University of Technology and Economics, Műegyetem rkp. 3., H-1111 Budapest, Hungary}
\affiliation{HUN-REN--BME Quantum Dynamics and Correlations Research Group, Műegyetem rkp. 3., H-1111 Budapest, Hungary}

\author{Viktor K\"{o}nye}
\affiliation{Institute for Theoretical Solid State Physics, IFW Dresden and W\"{u}rzburg-Dresden Cluster of Excellence ct.qmat, Helmholtzstr. 20, 01069 Dresden, Germany}
\affiliation{Institute for Theoretical Physics Amsterdam, University of Amsterdam,
Science Park 904, 1098 XH Amsterdam, The Netherlands}
\date{\today}

\begin{abstract}
Changes in the number of Weyl nodes in Weyl semimetals occur through merging processes, usually involving a pair of oppositely charged nodes.
More complicated processes involving multiple Weyl nodes are also possible, but they typically require fine tuning and are thus less stable.
In this work, we study how symmetries affect the allowed merging processes and their stability,
focusing on the combination of a two-fold rotation and time-reversal ($C_2\mathcal{T}$) symmetry. 
We find that, counter-intuitively, processes involving a merging of three nodes are more generic than processes involving only two nodes.
Our work suggests that multi-Weyl-merging may be observed in a large variety of quantum materials, and we discuss \ce{SrSi_2} and bilayer graphene as potential candidates.
\end{abstract}

\maketitle

\textcolor{blue}{\it Introduction ---}
Weyl semimetals are a class of gapless topological materials with robust two-fold band crossing points called \textit{Weyl points} (or nodes) in their Brillouin zone (BZ) \cite{Nielsen1981, Murakami2007, Wan2011, Burkov2011, Burkov2011a, Chiu2016, Yan2017, Armitage2018, Bernevig2018}.  
Weyl points are characterized by a linear low energy dispersion relation, and have $\pm 1$ \textit{topological charge} (often referred to as chirality) given by the value of the Chern number calculated for a surface enclosing them.

Since the topological charge in a region of momentum space with gapped spectrum on its boundary is a conserved quantity, isolated Weyl points cannot be gapped, but can only be moved or tilted by a small perturbation \cite{Xu2015, Soluyanov2015}. 
For larger perturbations though, it is possible to annihilate two Weyl points with opposite charges \cite{Volovik2018}. 
The phenomenon of pairwise creation and annihilation has been thoroughly investigated in the presence of various perturbations, including disorder \cite{Slager2017, Roy2018, pixley2016rare, Pixley2021}, magnetic fields, changes in magnetization  \cite{Zhang2017, Cano2017, Ghimire2019, Ray2022}, strain \cite{Sie2019, Liu2014, Slager2016, Facio2018}, phonons \cite{Singh2020} or high-frequency illumination \cite{Fu2017, Fu2019}.

Motivated by the prediction of merging processes involving three Weyl nodes (two positive and one negative)~\cite{Konye2021,Guba2023} in \ce{MoTe_2}~\cite{Tamai2016}, we are interested in studying the general stability conditions of various merging processes in the presence of $C_2\mathcal{T}$ symmetry -- the combination between a two-fold rotation ($C_2$) and time-reversal ($\mathcal{T}$) symmetry.
Thus, we expect our work will be relevant to a large class of quantum materials, including time-reversal invariant Weyl semimetals which host at least one two-fold rotation axis, but also magnetic systems in which $C_2\mathcal{T}$ symmetry is preserved even though $C_2$ and $\mathcal{T}$ are individually broken.

Adding symmetry constrains to conventional Weyl semimetals not only stabilizes higher-charge multi-Weyl points~\cite{Fang2012}, but it also restricts the possible positions and charges of the Weyl points in the BZ.
With $C_2\mathcal{T}$ symmetry, the configuration of Weyl points in the BZ is symmetric with respect to a $C_2\mathcal{T}$-invariant plane normal to the $C_2$ axis, which we take to be the $k_y=0$ momentum plane. The symmetry-related Weyl nodes at momenta $\vb{k} = (k_x, \pm k_y, k_z)$ have the same topological charge, and unpaired Weyl points are only allowed in the invariant plane and can only leave the plane by meeting another node of the same charge forming a pair with zero Euler class \cite{Ahn2019, Bouhon2020a}

\begin{figure}[t]
\centering
\includegraphics[width=8.6cm]{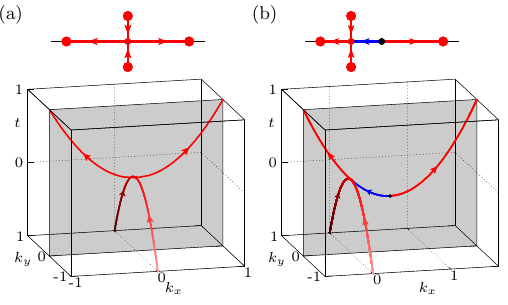} 
\caption{\label{fig:merging} Two type of allowed processes with $C_2\mathcal{T}$ symmetry in the $k_x-k_y$ plane with one control-parameter $t$. The $C_2\mathcal{T}$ invariant plane is shown in gray. The red/blue colors represent positive/negative charges of the Weyl nodes. 
The black dot represents pair-creation of nodes. 
The projection of the processes in the $k_x-k_y$ plane are shown above the graphs. Panel (a): the fine-tuned two-node process. Panel (b): the generic three-node process.}
\end{figure}

Consider processes in which the position of Weyl cones changes as a function of some control parameter, $t$. A pair of same-charge Weyl nodes start outside the $C_{2}\mathcal{T}$-invariant plane, and end as two (symmetry-unrelated) nodes in the plane.
One way such a process can occur is by moving the two nodes into the plane, forming a charge-two double-Weyl node, which then splits into two Weyl nodes within the plane.
This we call the \emph{two-node process}, see Fig.~\ref{fig:merging}(a).
There is, however, a different option: Before the nodes reach the $k_y=0$ plane, a pair creation occurs in the plane.
The two original nodes merge with the oppositely charged member of the created pair, forming a charge 1 cubic Weyl point, which then becomes one in-plane Weyl point, while the other final node is the other member of the pair.
This we call the \emph{three-node process}, see Fig.~\ref{fig:merging}(b).

We will show that, contrary to the naive expectation, the three-node process is stable, while the two-node process is not, explaining why only the former was observed in the numerical studies of \ce{MoTe_2}.
First we present an intuitive pictorial argument through Fig.~\ref{fig:merging}.
The processes are represented by ``Weyl world-lines'' in the $(t, k_x, k_y)$ space (restricting all nodes to $k_z=0$ without loss of generality) by two parabolas that touch at a single point where the nodes enter the plane.
In order to get a two-node process the parabolas must touch at their extremal points, while any other touching results in a three-node process, meaning that the two-node process requires fine tuning.

In this work we employ the mathematical framework of \emph{contact equivalence}~\cite{mond2020,Teramoto2017} 
to prove the above statement about the stability of merging processes with $C_2\mathcal{T}$ symmetry.
Stable processes are of experimental interest, as they can be accessed by only varying a single control parameter.
Then, we consider whether it is possible to observe both types of processes in the same system, without fine-tuning.
The answer lies in adding further symmetries, which make the two-node process stable, and can be broken by perturbations to produce the three-node process.
We showcase two physical examples, using simplified tight-binding models of \ce{SrSi_2} and bilayer graphene.

\textcolor{blue}{\it Setting up the problem ---}
In the following, we refer to the crystal quasimomenta $\vb{k}$ as \textit{configurational parameters} and call any further parameters such as magnetic field or strain as \textit{control parameters}. In our approach, we treat perturbations as the change of control parameters. 
The phase diagram divides the control parameter space into different regions, distinguished by the number and charges of Weyl points.
Generically, the borders between different regions correspond to pairwise creation and annihilation processes
where an unstable \textit{quadratic Weyl point} is formed which has zero topological charge (neutral) and its dispersion is quadratic in one direction.
This type of \emph{Weyl-point merging process} (a one-parameter family of configurations with non-constant number of nodes \footnote{We consider the two-node process also a merging process, even though the starting and final configuration has the same number of Weyl-nodes. This is consistent with our definition, because at a single point during the deformation, the number of nodes drops to 1, making it non-constant.}) is \emph{stable}, as small perturbations to the path in control parameter space do not change which configurations appear along the path.

To study the stability of merging processes we consider the most generic traceless effective $2\times 2$ Hamiltonians
$H(\vb{k}, \vb{t})=\vb*{\sigma} \cdot \vb{d}(\vb{k}, \vb{t})$,
where $\sigma_i$ are the Pauli matrices and $\vb{d}(\vb{k}, \vb{t})$ is a smooth $\mathbb{R}^3 \times \mathbb{R}^p \to \mathbb{R}^3$ function of three-dimensional momenta $\vb{k}$ and $p$ control parameters $\vb{t}$.
For fixed parameter values, we use the notation $\vb{d}_{\vb{t}}(\vb{k})$. The degeneracy points are at momenta where $\vb{d}_{\vb{t}}(\vb{k})=\vb{0}$.
We only consider processes that are local in $\vb{k}$-space and do not involve far away nodes.
More precisely, we only consider a compact $\vb{k}$-volume in momentum space and a compact $\vb{t}$-volume in the control parameter space around $\vb{t}=\vb{0}$, such that the spectrum of $H(\vb{k},\vb{t})$ is non-degenerate for momenta at the boundary of the $\vb{k}$-volume for any $\vb{t}$ taken from the $\vb{t}$-volume considered.

To formalize the intuitive definition, that Hamiltonians are qualitatively the same when they only differ, for example, in the locations and velocities of the Weyl nodes, we introduce the concept of \emph{contact equivalence} (or $\mathcal{K}$-equivalence) \cite{mond2020}.
Loosely speaking, we consider Hamiltonians equivalent if they can be transformed into each other by a smooth local reparametrization of $\vb k$-space and a smooth local $\vb k$-dependent reparametrization of the $\vb{d}$-space (inducing a deformation of the Hamiltonian that does not introduce new gap closings).
Introducing this equivalence relation has the advantage that it reduces the infinite-dimensional space of smooth functions $\vb{d}(\vb{k})$ to a function space which is finite-dimensional in our examples.
For precise definitions see App. \ref{app:def} and the Supplemental Material (SM), Sec.~I \cite{SM}.

A useful concept we will use is the \emph{codimension} of degeneracy points with respect to the contact equivalence.
Intuitively, it gives the number of parameters that have to be fine-tuned in order to find a given degeneracy point. 
A codimension 0 degeneracy point is called \emph{stable}. The stable degeneracy points in this sense are exactly the Weyl points.
For a higher, but finite codimension degeneracy point, it is possible to construct a \emph{versal unfolding} $\vb{d}(\vb{k}, \vb{t})$ of  $\vb{d}_0(\vb{k}) = \vb{d}(\vb{k}, \vb{0})$, where the space of control parameters $\vb{t}$ has the same dimension as the codimension of $\vb{d}_0(\vb{k})$ (for details see App. \ref{app:ver}).
The family of degeneracy point configurations given by $\vb{d}_{\vb{t}}(\vb{k})$ covers all the possible perturbations of $\vb{d}_0(\vb{k})$ up to contact equivalence, enabling us to make highly general statements using models with a finite number of parameters.
We refer to a specific choice of the versal unfolding $\vb{d}(\vb{k}, \vb{t})$ as a \emph{normal form} of the degeneracy point.

A single Weyl node, for example, has the normal form $\vb{d}(\vb{k}) = (k_x, k_y, k_z)$, where no tuning parameters appear, it being a stable (codimension 0) configuration.
As a second example, consider the neutral quadratic Weyl node, with normal form $\vb{d}(\vb{k}, t) = (k_x^2 + t, k_y, k_z)$.
The degeneracy point of $\vb{d}_0(\vb{k})$ has codimension 1, and can be accessed by tuning only one parameter.
For $t>0$ a gap opens, while for $t<0$ the degeneracy point splits into two oppositely charged Weyl nodes, realizing the creation/annihilation process.

We call a merging process \emph{stable}, if it connects two stable configurations, and does not pass through any configurations with codimension greater than 1, this guarantees that a small perturbation to the path in control parameter space does not change which configurations appear along the path.

\textcolor{blue}{\it Stability of merging processes with $C_2\mathcal{T}$ symmetry---}
Symmetries are accommodated in the above formalism by restricting to the space of $\vb{d}(\vb{k}, \vb{t})$ functions that obey the given symmetry constraints, and correspondingly restricting $\mathcal{K}$-equivalence transformations to those that maintain the symmetry (see App. \ref{app:symm}).
By introducing symmetries in the system we can change the codimension of certain degeneracy points, and consequently, the stability of certain processes.
In this manuscript we mainly focus on $C_2\mathcal{T}$ symmetry but many of the arguments can be applied 
to other types of symmetries.
As mentioned above, we consider $C_{2}$ rotations around the $k_y$ axis, which when combined with time reversal symmetry transforms the momentum space as $(k_x,k_y,k_z)\to (k_x,-k_y,k_z)$. The $C_{2}\mathcal{T}$ operator can be chosen as the complex conjugation
\footnote{Here we only consider $(C_{2}\mathcal{T})^2 = +1$, which is the case both for ordinary integer and half-integer spin particles. While it is possible to have situations where $(C_{2}\mathcal{T})^2 = -1$, this case would result in a Kramers degeneracy on the entire invariant plane, hence it would not give rise to any pointlike degeneracies.}
: this restricts the $\vb{d}$ vector to obey $d_{x/z}(k_x,k_y,k_z)=d_{x/z}(k_x,-k_y,k_z)$ and $d_{y}(k_x,k_y,k_z)=-d_{y}(k_x,-k_y,k_z)$.

\begin{figure}[t]
	\begin{center}
		\includegraphics[width=6cm]{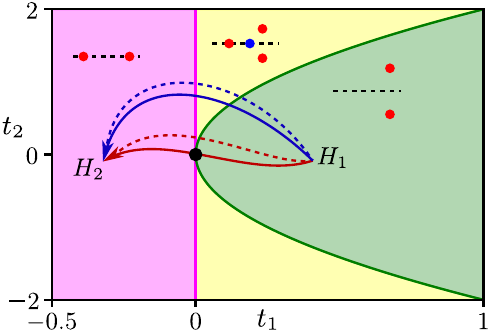}
	\end{center}
	\caption{Weyl node configurations of the perturbed double Weyl node~\eqref{eq:doubleWP}. The node is in the origin (black), and the perturbations ($t_1$, $t_2$) break it into three types of Weyl node configurations: A symmetric pair of Weyl nodes (green), 2 Weyl nodes on the symmetry plane (magenta), and 4 Weyl points (yellow). The regions are separated by the generic boundary of Weyl point annihilation on the symmetry plane (dark green) and the 3-point process (dark magenta). The red path shows a finetuned two-node process which is unstable, meaning that a small perturbation (dashed line) leads to a different process, whereas the blue path corresponds to a stable transition which happens through the 4-point configuration involving a pair creation on the symmetry plane and a three-node process.
	\label{fig:Weylbreaking}}
\end{figure}

To show the instability of the two-node process rigorously, we construct the $C_2\mathcal{T}$-symmetric versal unfolding of the most generic double-Weyl node and its perturbations, with the normal form 
\begin{equation}\label{eq:doubleWP}
\vb{d}(\vb{k},\vb{t})=
\begin{pmatrix}
k_x^2-k_y^2\\
k_x k_y\\
k_z
\end{pmatrix}+
\begin{pmatrix}
t_1 + t_2 k_x \\
0\\
0
\end{pmatrix}.
\end{equation}
This result shows that the double-Weyl point has codimension 2  in the presence of $C_2\mathcal{T}$ symmetry, making any process through this configuration unstable.
We emphasize that the low-energy Hamiltonian of any double-Weyl point subjected to any perturbation can be brought to the form \eqref{eq:doubleWP} with appropriate choice of parametrization; in this sense our results are generally applicable to any physical system with $C_2\mathcal{T}$ symmetry.

We find three distinct regions with stable degeneracy point configurations and two lines of codimension 1 configurations separating them in the control parameter space, shown in Fig.~\ref{fig:Weylbreaking} (for details see SM \cite{SM}, Sec.~III-IV).
We already saw that the pair creation process is stable (a statement not affected by the symmetry), and we find that the three-node process also becomes stable in the presence of the symmetry [the charge 1 cubic Weyl node has normal form $\vb{d}(\vb{k}, t) = (k_x, k_y^3 + t k_y, k_z)$].
In order to get the two-node process, the system must go through a single point corresponding to the double-Weyl node, any stable deformation results in a three-node process. 

\textcolor{blue}{\it Merging processes in $\ce{SrSi_2}$ ---} The three-node process has been theoretically predicted in the $C_2\mathcal{T}$ symmetric Weyl semimetal \ce{MoTe_2} \cite{Konye2021}.
As we argued above, the two-node process is unstable, explaining why only the three-node process was observed in a realistic model.
This raises the question, whether it is possible to observe the two-node process without fine-tuning parameters.

We propose the Weyl semimetal candidate \ce{SrSi_2} as a potential material in which to observe both three-node and two-node processes, the latter made possible without fine-tuning using the additional symmetries of the system.
\ce{SrSi_2} is a proposed double Weyl semimetal \cite{Huang2016, Singh2018} (with topological charge $2$) protected by fourfold rotation symmetry~\cite{Fang2016}.
Recent experimental observations \cite{yao2023observation} provided evidence that even though \ce{SrSi_2} is a trivial semimetal at ambient pressure, the predicted Weyl nodes can be observed by applying an external pressure to the material.
In the presence of $C_{4z}$ symmetry (four-fold rotation around the $k_z$ axis) the double-Weyl node is stable, while deformations that break $C_{4z}$, but preserve $C_{2z}$ and $C_{2y}\mathcal{T}$ stabilize the two-node process.
Further breaking the symmetries down to $C_{2y}\mathcal{T}$ it is possible to observe the three-node process.

We construct a simple model that reproduces the key features of \ce{SrSi_2}, such that it has chiral cubic and time-reversal symmetry and hosts double-Weyl nodes on the fourfold axes.
By looking at the explicit form of the $C_{2y}\mathcal{T}$ and $C_{4z}$ operators acting on Eq.~\eqref{eq:doubleWP} (see SM \cite{SM}, Sec.~V) we conclude that the minimal model needs local degrees of freedom with a total angular momentum of at least $J = 3/2$.
We use \texttt{qsymm} \cite{Varjas_2018} to get the most general 4x4 $\vb{k}\cdot\vb{p}$ model to quadratic order with chiral cubic and time-reversal symmetry, resulting in:
\begin{equation}
\label{eq:qsymmkdotp}
\begin{split}
H(\vb{k})=&\left(c_0 + c_4 \vb{k}^2\right) \mathbbm{1} + c_1 \vb{k}\cdot\vb{J} + c_2 \vb{k}\cdot\vb{S} \\
& + c_3 \sum_{i\neq j}\left(k_i k_j J_i J_j\right) + c_5 \sum_i k_i^2 J_i^2,
\end{split}
\end{equation}
where $\vb{J}$ is the vector of spin-$3/2$ angular momentum matrices, and $\vb{S}$ is a vector of effective spin-$1/2$ angular momentum matrices given by $S_i = \frac{7}{6} J_i - \frac{2}{3} J_i^3$.
This model has double Weyl nodes at $\vb{k}= (0,0,\pm c_1/c_5)$ and symmetry related points.

In order to split the double Weyl node we add a magnetic field $B_x$ and uniaxial strain $s$ in the $x$ direction, $H_{\text{pert}} = B_x J_x + s J_x^2$.
The last term arises as a crystal field term changing the energy of $p_x$ orbitals compared to $p_{y/z}$ restricted to the $J=3/2$ subspace of a spin-orbit coupled atom.
The uniaxial strain breaks the $C_{4z}$ symmetry down to $C_{2z}$, and
adding the magnetic field in the $x$ direction breaks $C_{2z}$ and only preserves $C_{2y}\mathcal{T}$.

\begin{figure}[h]
	\begin{center}
		\includegraphics[width=8.6cm]{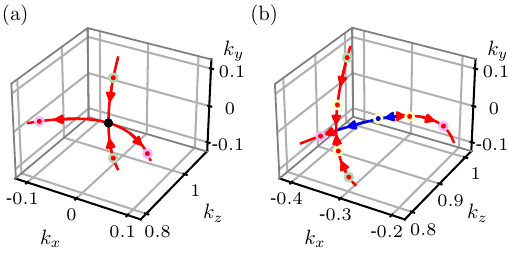}
	\end{center}
	\caption{Weyl node processes under varying uniaxial strain $s$ and magnetic field $B_x$ in the Eq.~\eqref{eq:qsymmkdotp} model for \ce{SrSi_2}.
    The black point has charge $+2$. Green, yellow and magenta indicate the corresponding phases of Fig.~\ref{fig:Weylbreaking}.
    Model parameters are $c_0=0$, $c_2=-0.6$, $c_3=0.75$, $c_4=1$, $c_5=-1$ (in units of $c_1$). Strain is taken in the range $s\in [-0.1,0.1]$, the arrows indicate the Weyl node movement direction for increasing $s$. (a)  Two-node process with $B_x=0$. (b) Pair creation and three-node process with $B_x=0.5$. The Weyl node configurations are obtained numerically from a tight-binding model on a cubic lattice constructed from Eq.~\eqref{eq:qsymmkdotp} (see SM \cite{SM}, Sec.~V).
	\label{fig:merging_SrSi}}
\end{figure}

The different processes under varying strain and magnetic field are shown in Fig.~\ref{fig:merging_SrSi}.
Without a magnetic field setting $s<0$  splits the double Weyl node creating two nodes inside the $C_{2y}\mathcal{T}$ symmetry plane while setting $s>0$ creates two nodes outside the plane. In this case the additional symmetry constraints due to $C_{2z}$ make the observation of the unstable two-node process possible.
In the presence of a magnetic field $B_x$, the $C_{2z}$ symmetry is broken. The system goes from having two out of plane nodes to two in-plane nodes through a pair creation and a three-node process.

In Sec.~V of the SM \cite{SM} we demonstrate explicitly that the minimal model Eq.~\eqref{eq:doubleWP} is recovered from the 4-band model using quasi-degenerate perturbation theory.

\textcolor{blue}{\it Merging processes in bilayer graphene ---} As a second example, we show the possibility of observing two- and three-node merging processes with the Dirac cones of bilayer graphene, that are analogous to Weyl cones in Weyl semimetals.

Bilayer graphene \cite{Novoselov2006, McCann2006,  McCann_2013}
is formed through an $A$-$B$ stacking of two graphene layers.
When only intralayer nearest neighbor and vertical interlayer hoppings are considered there are two parabolic degenerate points at $K$ and $K'$ in the BZ.
The two degenerate points are protected by chiral symmetry and have opposite topological charges $\pm2$, given by the winding number calculated on a closed loop around them \cite{Park2011}. 

Adding a skew interlayer hopping term in the Hamiltonian leads to a Lifshitz transition and splits the original degenarate point at $K$ with charge $+2$ into four Dirac cones: three with charges $1$ and one with $-1$ \cite{McCann_2013, Volovik2018}.
Adding uniaxial strain in the $x$ direction splits the charge $2$ degeneracy in two charge $1$ Dirac cones \cite{Mucha2011, Mucha2011_2,McCann_2013, Varlet2015}.

The linearized Hamiltonian around the $K$ point in the presence of skew hoppings and strain reads \cite{McCann_2013, Varlet2015}
\begin{equation}
\label{eq:graphenelowenergy}
H(\vb{k})=\begin{pmatrix}

0           & v\pi^\dag & 0       & v_3\pi +w\\
v\pi        & 0         &\gamma_1 & 0 \\
0           &\gamma_1   & 0       & v\pi^\dag \\
v_3\pi^\dag +w& 0         &v\pi     & 0

\end{pmatrix},
\end{equation}
where $\pi=k_x+ik_y$,  $v$ is the intralayer band velocity, $\gamma_1$ is the vertical interlayer hopping, $v_3$ is the skew velocity and $w$ is the strain parameter.

This system has $M_y\mathcal{T}$ symmetry (mirror symmetry with respect to the $k_y$ axis times time-reversal symmetry), which is analogous to the $C_2\mathcal{T}$ symmetry in the three-dimensional case. In this case Dirac cones outside the $k_y=0$ axis come in pairs and are related to each other by the $M_y\mathcal{T}$ symmetry and have the same charge.

We consider the control parameters to be the $v_3$ skew velocity and $w$ strain.
$v_3$ can be independently tuned from the uniaxial strain by applying external pressure to the layers.
To understand which merging processes can occur as we tune $w$ and $v_3$, we derive a system of equations similar to Eq.~\eqref{eq:doubleWP} (see SM \cite{SM}, Sec.~VI). 
In the case where $v_3=0$, the two Dirac points at $k_y\neq0$ enter the symmetry axis through the unstable two-node process.
When $v_3\neq 0$, the only way for bringing the two Dirac cones outside the axis into the symmetry axis is through pair creation of Dirac cones and a three-node process, similar to the three-dimensional case. 

\textcolor{blue}{\it Conclusion ---}
We have shown that merging processes of multiple Weyl nodes can be observed in Weyl semimetals with $C_2\mathcal{T}$ symmetry.
Counter-intuitively, the simultaneous merging of three Weyl nodes is generic, while two-node processes and the formation of double-Weyl nodes are unstable, since they require either additional symmetry constraints or fine tuning of the Hamiltonian parameters. 

We used the Weyl semimetal candidate \ce{SrSi_2} as a case study. 
By using an effective tight-binding model we have shown that it is possible to reduce the symmetry of the system to $C_2\mathcal{T}$ and to observe a stable three-node process using strain and magnetic field. 
Since $C_2\mathcal{T}$-symmetric Weyl semimetals are ubiquitous in the literature \cite{Tamai2016, Xu2018, Koepernik2016, Weng2016, Chen2022, Chen2023}, it could be possible to observe the same type of merging in other materials under the right conditions.
For example, a two Weyl node merging process was predicted in ReO$_3$ \cite{Chen2023} in the presence of strain.
By adding perturbations which lower the three $C_2\mathcal{T}$ symmetries to only one, it could be possible to observe triple Weyl merging processes. Furthermore, since the three node process requires only a local $C_2\mathcal{T}$ symmetry, similar merging processes could be observed in systems with a broader set of symmetries away from high symmetry points. 

We have shown that similar processes can be realized in two-dimensional chiral Dirac semimetals in the presence of $M\mathcal{T}$ symmetry, since the symmetry constraints are analogous to the three-dimensional case. 
This can be seen by applying strain to bilayer graphene \cite{Mucha2011, Mucha2011_2, Son2011, Mariani2012, Varlet2015}. In the presence of a skew hopping and strain, the Dirac cones will generically enter or exit the $M\mathcal{T}$ symmetry axis by creating a new pair of Dirac cones and undergoing a stable three-point process.

Our results also have applications beyond the study of topologically-protected band degeneracies in semiemtals. 
For instance, they can be used to explore merging processes between analog Weyl points in meta-materials, such as interacting spin systems \cite{Scherbl2019, Frank2020}, multiterminal Josephson junctions \cite{vanHeck2014, Riwar2016} and spring systems \cite{Guba2023}, and between other topological points, e.g magnetic Bloch points in spin systems \cite{Tapia2024, Kuchkin2024}.

In this paper, only the presence of a single $C_2\mathcal{T}$ symmetry was considered. The formalism presented here can be applied to study other lattice symmetries, providing a foundation for exploring merging processes involving multiple Weyl nodes.

\begin{acknowledgments}
{\it Author contributions ---}
The initial project idea was formulated by V.K. after a discussion with A.P. about the stability of the three-node process in MoTe$_2$.
A.P. had the idea that the stability of the three-node process can be explained using codimensions.
Gy.F. performed the stability analysis and confirmed this.
G.P. suggested using contact equivalence, and explained this theory to the others. Gy.F., G.P. and D.V. formulated the contact equivalence analysis.
I.C.F. suggested the potential material realizations of \ce{SrSi_2}, and, in parallel with Gy.F., of bilayer graphene.
D.V. constructed the 4-band model of \ce{SrSi_2} and performed the perturbation theory calculation with input from G.N..
G.N. constructed the tight binding model of \ce{SrSi_2} 
from the 4-band model given by D.V. 
and carried out the numerical calculations needed to produce Fig.~\ref{fig:merging_SrSi} with input from V.K.. 
G.N. has also carried out the analytical calculation and built the phase diagram for bilayer graphene  and \ce{SrSi_2} with input from V.K., I.C.F and Gy.F..
G.N. and V.K. wrote the initial draft of the manuscript.
A.P. and V.K. supervised the project. 
All authors took part in interpreting the results and writing the final version of the manuscript.

{\it Acknowledgments ---}
We thank J{\'a}nos Asb{\'o}th for fruitful discussions. Gy.F., G.P., and A.P. was supported by the Ministry of Culture and Innovation and the National Research, Development and Innovation Office within the Quantum Information National Laboratory of Hungary (Grant No. 2022-2.1.1-NL-2022-00004), and the National Research, Development and Innovation Office via the OTKA Grant No. 132146.
A.P. received funding from the HUN-REN Hungarian Research Network.
G.N., D.V., I.C.F, and V.K. acknowledge support from the Deutsche Forschungsgemeinschaft (DFG, German Research Foundation) under Germany’s Excellence Strategy through the W\"{u}rzburg-Dresden Cluster of Excellence on Complexity and
Topology in Quantum Matter – ct.qmat (EXC 2147, project-ids 390858490 and 392019).
D.V. received funding from the National Research, Development and Innovation Office of Hungary under OTKA grant no. FK 146499.
 
\end{acknowledgments}

\bibliography{references}

\newpage
\appendix

\section*{End Matter}
Here we provide additional, specialized information, including the precise definitions and essential mathematical details of the contact equivalence formalism that we use.
We relegate illustrative examples and further details to the SM~\cite{SM}.
In order to keep the SM self-contained, we repeat the following sections at the appropriate locations in the SM with minor modifications.

\section{Definitions for contact equivalence of degeneracy points}\label{app:def}

We treat isolated two-fold degeneracy points in three-dimensional band-structures.
We consider traceless $2\times 2$ Hamiltonians, as any Hamiltonian can be smoothly deformed into a traceless one by introducing a $\vb{k}$-dependent energy shift.
The restriction to $2\times 2$ Hamiltonians is without loss of generality: Starting from a larger Hamiltonian that has a twofold degeneracy point, we can perform quasi-degenerate perturbation theory restricted to the two bands involved in the crossing, and construct an effective $2\times 2$ Hamiltonian in this subspace.
As long as these two bands are separated from all other bands by a finite gap in some vicinity of the degeneracy point, small perturbations of the full Hamiltonian change the effective Hamiltonian continuously, and cannot create degeneracies between these bands and any others.

Hence the degeneracy points are studied by writing
\begin{equation}
    H(\vb k) = \vb*{\sigma} \cdot \vb{d}(\vb{k}),
\end{equation}
and analyzing the $\vb{d}:\mathbb{R}^3 \to \mathbb{R}^3$ function.
As we are only interested in properties in the vicinity of the degeneracy point, which we fix to be at $\vb{k=0}$, we treat $\vb d$ as a \emph{map germ}, i.e. an equivalence class of functions that are equal in an open neighborhood of $\vb{0}$ (for example a ball $|\vb{k}| < \epsilon$ with fixed $\epsilon$).
We denote by $\vb{d}: (\mathbb{R}^3, \vb{0}) \to (\mathbb{R}^3, \vb{d}(\vb{0}))$ the germ $\vb{d}$ at $\vb{0}$.
Since we consider degerenacy points, we only consider germs with $\vb{d}(\vb{0})=\vb{0}$.

We call two Hamiltonians described by the map germs $\vb{d}(\vb{k})$ and $\vb{d}'(\vb{k})$ contact equivalent ($\mathcal{K}$-equivalent), if there exists a germ of diffeomorphism $\Psi: (\mathbb{R}^3 \times \mathbb{R}^3, 0) \to (\mathbb{R}^3 \times \mathbb{R}^3, 0)$ in the special form $\Psi(\vb{k}, \vb{d})= (\vb*{\phi}(\vb{k}), \vb*{\psi}(\vb{k}, \vb{d}))$ such that $\vb*{\phi}: (\mathbb{R}^3, 0) \to (\mathbb{R}^3, 0)$ is a germ of diffeomorphism with $\vb*{\phi}(\vb{0})=\vb{0}$, and $\vb*{\psi}(\vb{k}, \cdot): \mathbb{R}^3 \to \mathbb{R}^3$ are locally defined  $\vb{k}$-dependent diffeomorphisms with $\vb*{\psi}(\vb{k}, \vb{0})=\vb{0}$, and
\begin{equation}
    \vb{d}'(\vb{k}) = \vb*{\psi}\left(\vb{k}, \vb{d}\!\left(\vb*{\phi}\left(\vb{k}\right)\right)\right).
\end{equation}
Intuitively,
$\vb*{\phi}$ is a smooth and invertible reparametrization of $\vb k$-space, while $\vb*{\psi}$ is a smooth and invertible, $\vb k$-dependent reparametrization of the space of Hamiltonians ($\vb d$-space).
We get equivalent results if we restrict to $\vb*{\psi}$'s that are linear in their second argument~\cite[Prop. 4.2.]{mond2020}:
\begin{equation}
    \vb*{\psi}(\vb{k}, \vb{d}) = A(\vb{k}) \vb{d},
\end{equation}
where $A$ is a smooth map germ taking values from invertible $3\times 3$ matrices, $A: \mathbb{R}^3 \to \textnormal{GL}(3, \mathbb{R})$.
(Note that $\vb{d}$ without an explicit $\vb{k}$-dependence simply denotes a variable in $\mathbb{R}^3$, not a map germ.)
With this, the contact equivalence condition reads
\begin{equation}
\label{eqn:KLeq_trf}
    \vb{d}'(\vb{k}) = A(\vb{k}) \vb{d}\!\left(\vb*{\phi}\left(\vb{k}\right)\right).
\end{equation}

Following \cite[Lemma 4.1.]{mond2020} (see also Ref.~\onlinecite{Teramoto2017}) we investigate infinitesimal $\mathcal{K}$-equivalence transformations, and construct the tangent space corresponding to transformations that do not leave the orbit of $\mathcal{K}$-equivalence.
Consider a one-parameter family of transformations given by $A_t(\vb{k})$ and $\vb*{\phi}_t(\vb{k})$ where $t \in \mathbb{R}$, and $A_0(\vb{k}) = \mathbbm{1}$ and $\vb*{\phi}_0(\vb{k}) = \vb{k}$ corresponding to the identity transformation.
Note that by definition $\vb*{\phi}_t$ fixes the origin, i.e., $\vb*{\phi}_t(\vb{0})=\vb{0}$ holds for every $t$.
An infinitesimal transformation on $\vb{d}(\vb{k})$ reads
\begin{align}
\label{eqn:Ktrf}
    \left.\frac{d}{dt}\vb{d}_t(\vb{k})\right|_{t=0} &=
    \left.\frac{d}{dt}\left[A_t(\vb{k})\vb{d}_0\!\left(\vb*{\phi}_t\left(\vb{k}\right)\right)\right]\right|_{t=0} \\
    &= B(\vb{k}) \vb{d}_0(\vb{k}) + J(\vb{k}) \vb{c}(\vb{k}),
\end{align}
where
\begin{eqnarray}
    B(\vb{k}) &=& \left.\frac{d}{dt}A_t(\vb{k})\right|_{t=0},\nonumber\\
    \vb{c}(\vb{k}) &=& \left.\frac{d}{dt}\vb*{\phi}_t(\vb{k})\right|_{t=0}, \\
    J(\vb{k})_{ij} &=& \frac{\partial d_{0i}}{\partial k_j}(\vb{k}).\nonumber
\end{eqnarray}
The functions $B(\vb{k})$ and $\vb{c}(\vb{k})$ are arbitrary matrix and vector-valued smooth functions (power series) with $\vb{c}(\vb{0})=\vb{0}$ and $J(\vb{k})$ is the Jacobian of $\vb d(\vb k)$.
All the infinitesimal transformations that can be written in the form of \eqref{eqn:Ktrf} remain at first order in the orbit of $\mathcal{K}$-equivalence, forming the $\mathcal{K}$-tangent space.
Power series that cannot be represented as \eqref{eqn:Ktrf} form \emph{transverse perturbations} of $\vb{d}_0$, which leave the orbit of $\mathcal{K}$-equivalence.
The space of transverse perturbations fixing the origin is formed by the quotient of the space of all power series with zero constant term, with the space of power series of the form \eqref{eqn:Ktrf}. The dimension of the quotient space is called the $\mathcal{K}$-codimension of the germ $\vb{d}(\vb{k})$.

However, we want to consider perturbations which possibly move the origin.
For it we need the extended tangent space ($\mathcal{K}_e$-tangent space), defined in the same way as the right side of \eqref{eqn:Ktrf}, but without the restriction $\vb{c}(\vb{0})=\vb{0}$.
Hence we allow for perturbed functions with $\vb{d}_{\vb{t}}(\vb{0}) \neq \vb{0}$ for $\vb{t}\neq \vb{0}$.
The space of all  perturbations transverse to the extended tangent space is formed by the quotient of the space of all power series (with possibly non-zero constant term) with the space of power series of the form \eqref{eqn:Ktrf} (possibly with $\vb{c}(\vb{0})\neq\vb{0}$).
The dimension of the quotient space is called the $\mathcal{K}_e$-codimension of the germ $\vb{d}(\vb{k})$, the subscript refers to `extended'.

We demonstrate the calculation of the $\mathcal{K}_e$-codimension and transverse perturbations using elementary techniques in Sec.~I of the SM~\cite{SM}.
In general this calculation can be carried out in the following way:
First we show that all vector valued monomials above a certain degree can be generated by a suitable choice of $B(\vb{k})$ and $\vb{c}(\vb{k})$.
In the next step, we restrict $B(\vb{k})$ and $\vb{c}(\vb{k})$ to be polynomials of this maximal order, restricting them to a finite dimensional subspace of the vector and matrix valued polynomial ring.
Then we evaluate \eqref{eqn:Ktrf} for a set of basis functions in this space, and expand the result (also restricted to this finite dimensional subspace) in a basis, resulting in a matrix equation between the free coefficients in $B(\vb{k})$ and $\vb{c}(\vb{k})$ and the coefficients of the resulting $\frac{d}{dt}\vb{d}_t(\vb{k})$.
Calculating the rank deficiency of this matrix (the dimension of its null space) gives the codimension, and a basis of the complement of its image yields the transverse perturbations.

\section{Unfoldings and versality}
\label{app:ver}

Perturbations or deformations of germs are formulated via the notion of \emph{unfoldings}. An unfolding of $\vb{d}(\vb{k})$ is a germ in form $\vb{D}(\vb{k}, \vb{t})=(\vb{d}(\vb{k}, \vb{t}), \vb{t})$, where $\vb{t}\in \mathbb{R}^n$ for some $n$ and $\vb{d}(\vb{k}, \vb{t}=\vb{0}) = \vb{d}(\vb{k})$. Its first component $\vb{d}(\vb{k}, \vb{t})$ is the perturbation or deformation corresponding to the control parameter $\vb{t}$.
Note that for a fixed $\vb{t}$ value, the perturbation $\vb{k} \mapsto \vb{d}(\vb{k}, \vb{t})$ is a map of $\vb{k}$, but not a germ. However, $\vb{D}(\vb{k}, \vb{t})$ is a germ thanks to the inclusion of the trivial $\vb{t}$ component.
We fix a representative map of the unfolding $\vb{D}(\vb{k}, \vb{t})$ with a sufficiently small neighborhood $|\vb{k}| < \epsilon$ and $|\vb{t}| < \delta$, such that $\vb{d}(\vb{k}, \vb{t})=0$ has no solutions with $|\vb{t}| <\delta $ and $|\vb{k}| = \epsilon$, i.e., the gap does not close on the boundary of the $\epsilon$-ball in the $\vb{k}$-space. This makes the analysis shown in Fig.~\ref{fig:Weylbreaking} in the main text precise.

An unfolding $\vb{D}_v\!\left(\vb{k}, \vb{t}\right)=(\vb{d}_v(\vb{k}, \vb{t}), \vb{t})$ ($\vb{t} \in \mathbb{R}^m$) is \emph{versal} (more precisely, $\mathcal{K}$-versal), if every other unfolding can be induced from it up to contact equivalence: given any unfolding $\vb{D}(\vb{k}, \vb{t})$ ($\vb{t} \in \mathbb{R}^n$), there exists a base-change function (pull-back) $\vb{h} : \mathbb{R}^n \to \mathbb{R}^m$ such that
\begin{equation}
    \widetilde{\vb{D}}(\vb{k}, \vb{t}) := (\vb{d}_v\!\left(\vb{k}, \vb{h}(\vb{t})), \vb{t}\right)
\end{equation}
agrees with $\vb{D}(\vb{k}, \vb{t})$ up to \emph{contact equivalence of unfoldings}, see Ref.~\onlinecite{Wall,Damon} for a precise definition. It is also \emph{miniversal}, if there exists no versal unfolding with a lower number of parameters.

By the \emph{unfolding theorem}~\cite{Wall, Damon}, a miniversal unfolding can be constructed as follows:
First we choose a basis for perturbations $\vb{e}_i(\vb{k})$ transverse to the $\mathcal{K}_e$-tangent space as described in the previous section.
The number of basis elements is the $\mathcal{K}_e$-codimension of $\vb{d}(\vb{k})$ by definition.
Adding the basis elements to $\vb{d}(\vb{k})$  with arbitrary parameters ($\vb{t} \in \mathbb{R}^n$ where $n$ equals the codimension), we get an unfolding:
\begin{equation}
    \vb{D}_v\!\left(\vb{k}, \vb{t}\right)=\left(\vb{d}(\vb{k}) + \sum_{i=1}^n t_i \vb{e}_i(\vb{k}), \vb{t}\right).
\end{equation}
The unfolding theorem states that this unfolding is miniversal.

\section{Symmetry constraints}
\label{app:symm}
We modify the above construction of the tangent space, transverse perturbations, and unfoldings to be applicable in the presence of symmetries, by restricting to the subspace of functions that respect the prescribed symmetries.

A general symmetry poses a restricion on $\vb{d}(\vb{k})$ of the form
\begin{equation}
    V \vb{d}(\vb{k}) = \vb{d}(R \vb{k}),
\end{equation}
where $V$ and $R$ are $3\times 3$ orthogonal matrices acting in $\vb{d}$ and $\vb{k}$ space respectively.
We demand that the contact equivalence transformation of Eq.~\eqref{eqn:KLeq_trf} preserves this symmetry, which results in the constraints on the symmetry-compatible transformations described by $A(\vb{k})$ and $\vb*{\phi}\!\left(\vb{k}\right)$:
\begin{align}
    V A(\vb{k}) =& A(R\vb{k}) V, \\
    R\, \vb*{\phi}\!\left(\vb{k}\right) =& \vb*{\phi}\!\left(R \vb{k}\right).
\end{align} 
Constraints of the same form also apply to $B(\vb{k})$ and $\vb{c}(\vb{k})$ respectively.
The equivalence defined by such symmetry-restricted transformations is called \emph{equivariant contact equivalence}.
The computation of the codimension and normal form follows the same procedure as without symmetry, except that the spaces of available functions are restricted.

Although in our examples these restrictions are straightforward (see Sec. II of the SM~\cite{SM}), for the general theory we refer to Ref.~\onlinecite{Roberts}, \onlinecite{WallJet}.
Note that this restriction guarantees that \emph{any} symmetric $\vb{d}(\vb{k})$ is mapped to a symmetric one by the allowed contact equivalence transformations.
There are cases where two symmetric $\vb{d}(\vb{k})$'s are connected by an unrestricted contact equivalence transformation, but not by a restricted one.
We give a non-trivial example of this situation in Sec. II of the SM~\cite{SM}.
In the specific case of of $C^y_2\mathcal{T}$ symmetry introduced in the main text, the function $\vb{d}(\vb{k})$ is restricted as:
\begin{align}
\label{eq:Hsymm}
d_x(k_x,-k_y,k_z)&= d_x(k_x,k_y,k_z)\nonumber\\
d_y(k_x,-k_y,k_z)&= -d_y(k_x,k_y,k_z)\\
d_z(k_x,-k_y,k_z)&= d_z(k_x,k_y,k_z)\nonumber.
\end{align}
In the notation above this corresponds to $V = R = \operatorname{diag}(1, -1, 1)$.

\end{document}


\title{Supplemental Material -- Stability of Weyl node merging processes under symmetry constraints}
\author{Gabriele Naselli}
\thanks{These two authors contributed equally}
\affiliation{Institute for Theoretical Solid State Physics, IFW Dresden and W\"{u}rzburg-Dresden Cluster of Excellence ct.qmat, Helmholtzstr. 20, 01069 Dresden, Germany}

\author{György Frank}
\thanks{These two authors contributed equally}
\affiliation{Department of Theoretical Physics, Institute of Physics, Budapest University of Technology and Economics, Műegyetem rkp. 3., H-1111 Budapest, Hungary}

\author{Dániel Varjas}
\affiliation{Institute for Theoretical Solid State Physics, IFW Dresden and W\"{u}rzburg-Dresden Cluster of Excellence ct.qmat, Helmholtzstr. 20, 01069 Dresden, Germany}
\affiliation{Department of Theoretical Physics, Institute of Physics, Budapest University of Technology and Economics, Műegyetem rkp. 3., H-1111 Budapest, Hungary}
\affiliation{Max Planck Institute for the Physics of Complex Systems, N\"{o}thnitzer Strasse 38, 01187 Dresden, Germany}

\author{Ion Cosma Fulga}
\affiliation{Institute for Theoretical Solid State Physics, IFW Dresden and W\"{u}rzburg-Dresden Cluster of Excellence ct.qmat, Helmholtzstr. 20, 01069 Dresden, Germany}

\author{Gergő Pintér}
\affiliation{Department of Theoretical Physics, Institute of Physics, Budapest University of Technology and Economics, Műegyetem rkp. 3., H-1111 Budapest, Hungary}

\author{András Pályi}
\affiliation{Department of Theoretical Physics, Institute of Physics, Budapest University of Technology and Economics, Műegyetem rkp. 3., H-1111 Budapest, Hungary}
\affiliation{HUN-REN--BME Quantum Dynamics and Correlations Research Group, Műegyetem rkp. 3., H-1111 Budapest, Hungary}

\author{Viktor K\"{o}nye}
\affiliation{Institute for Theoretical Solid State Physics, IFW Dresden and W\"{u}rzburg-Dresden Cluster of Excellence ct.qmat, Helmholtzstr. 20, 01069 Dresden, Germany}
\affiliation{Institute for Theoretical Physics Amsterdam, University of Amsterdam, Science Park904, 1098 XH Amsterdam, The Netherlands}
\date{\today}

\maketitle

We summarize the nomenclature we use in Table~\ref{tab:nomenclature}. In the rest of this Supplemental Material we present precise definitions and results of the codimension calculations, and details of the physical models studied in the manuscript.

\begin{table}[ht]
        \renewcommand*{\arraystretch}{1.4}
	\begin{tabular}{|l|p{7cm}|p{6cm}|}
	    \hline
		Term&Description&Remark \\
		\hline\hline
		Degeneracy point/node & {Point-like 2-fold band degeneracy.} & {Includes all types of nodes below.}\\
		\hline
   Topological charge & {The Chern number on a closed surface  enclosing the degeneracy.} & \\
   \hline
  Configurational parameters & {The set of the first 3 parameters where Weyl points are generic features} & {Typically associated with momenta $\vb{k}$.}\\
		\hline
    Control parameters & {The remaining parameters which control the motion and merging of Weyl points.} & {Typically denoted by $\vb{t}$.}\\
		\hline
  Weyl phase diagram & {The partitioning of the control parameter space into separate regions according to  corresponding number and type of the band degeneracies in the configurational parameter space.} & \\
  \hline
  Perturbation & {Small change of control parameters.} & \\
  \hline
  Stability of a degeneracy & {Robustness of a degeneracy type against perturbations.} & {To define degeneracy type one needs an equivalence relation of the degeneracies.}\\
  \hline
  Codimension & {Number of directions in control parameter space
  that lead out of the given degeneracy type.} &{Stable degeneracies have codimension 0.}\\
  \hline
  Process & {Motion and merging of degeneracies corresponding to a space curve in the control parameter space.}& {Perturbation of a process is a perturbation of the space curve with fixed endpoints.}\\
  \hline
  Stability of a process & {Robustness of the merging types in a process against perturbations.}& {Stable processes involve the configurations whose codimension is at most 1.}\\
  \hline
	\end{tabular}
	\caption{Summary of the nomenclature used in this paper.  \label{tab:nomenclature}}
\end{table}

\section{Definitions and mathematical details}

\subsection{Contact equivalence of degeneracy points}\label{ss:conteq}

We treat isolated two-fold degeneracy points in three-dimensional band-structures.
We consider traceless $2\times 2$ Hamiltonians, as any Hamiltonian can be smoothly deformed into a traceless one by introducing a $\vb{k}$-dependent energy shift.
The restriction to $2\times 2$ Hamiltonians is without loss of generality: Starting from a larger Hamiltonian that has a twofold degeneracy point, we can perform quasi-degenerate perturbation theory restricted to the two bands involved in the crossing, and construct an effective $2\times 2$ Hamiltonian in this subspace.
As long as these two bands are separated from all other bands by a finite gap in some vicinity of the degeneracy point, small perturbations of the full Hamiltonian change the effective Hamiltonian continuously, and cannot create degeneracies between these bands and any others.

Hence the degeneracy points are studied by writing
\begin{equation}
    H(\vb k) = \vb*{\sigma} \cdot \vb{d}(\vb{k}),
\end{equation}
and analyzing the $\vb{d}:\mathbb{R}^3 \to \mathbb{R}^3$ function.
As we are only interested in properties in the vicinity of the degeneracy point, which we fix to be at $\vb{k=0}$, we treat $\vb d$ as a \emph{map germ}, i.e. an equivalence class of functions that are equal in an open neighborhood of $\vb{0}$ (for example a ball $|\vb{k}| < \epsilon$ with fixed $\epsilon$).
We denote by $\vb{d}: (\mathbb{R}^3, \vb{0}) \to (\mathbb{R}^3, \vb{d}(\vb{0}))$ the germ $\vb{d}$ at $\vb{0}$.
Since we consider degerenacy points, we only consider germs with $\vb{d}(\vb{0})=\vb{0}$.

We call two Hamiltonians described by the map germs $\vb{d}(\vb{k})$ and $\vb{d}'(\vb{k})$ contact equivalent ($\mathcal{K}$-equivalent), if there exists a germ of diffeomorphism $\Psi: (\mathbb{R}^3 \times \mathbb{R}^3, 0) \to (\mathbb{R}^3 \times \mathbb{R}^3, 0)$ in the special form $\Psi(\vb{k}, \vb{d})= (\vb*{\phi}(\vb{k}), \vb*{\psi}(\vb{k}, \vb{d}))$ such that $\vb*{\phi}: (\mathbb{R}^3, 0) \to (\mathbb{R}^3, 0)$ is a germ of diffeomorphism with $\vb*{\phi}(\vb{0})=\vb{0}$, and $\vb*{\psi}(\vb{k}, \cdot): \mathbb{R}^3 \to \mathbb{R}^3$ are locally defined  $\vb{k}$-dependent diffeomorphisms with $\vb*{\psi}(\vb{k}, \vb{0})=\vb{0}$, and
\begin{equation}
    \vb{d}'(\vb{k}) = \vb*{\psi}\left(\vb{k}, \vb{d}\!\left(\vb*{\phi}\left(\vb{k}\right)\right)\right).
\end{equation}
Intuitively,
$\vb*{\phi}$ is a smooth and invertible reparametrization of $\vb k$-space, while $\vb*{\psi}$ is a smooth and invertible, $\vb k$-dependent reparametrization of the space of Hamiltonians ($\vb d$-space).
We get equivalent results if we restrict to $\vb*{\psi}$'s that are linear in their second argument~\cite[Prop. 4.2.]{mond2020}:
\begin{equation}
    \vb*{\psi}(\vb{k}, \vb{d}) = A(\vb{k}) \vb{d},
\end{equation}
where $A$ is a smooth map germ taking values from invertible $3\times 3$ matrices, $A: \mathbb{R}^3 \to \textnormal{GL}(3, \mathbb{R})$.
(Note that $\vb{d}$ without an explicit $\vb{k}$-dependence simply denotes a variable in $\mathbb{R}^3$, not a map germ.)
With this, the contact equivalence condition reads
\begin{equation}
\label{eqn:KLeq_trf}
    \vb{d}'(\vb{k}) = A(\vb{k}) \vb{d}\!\left(\vb*{\phi}\left(\vb{k}\right)\right).
\end{equation}

Following \cite[Lemma 4.1.]{mond2020} (see also Ref.~\onlinecite{Teramoto2017}) we investigate infinitesimal $\mathcal{K}$-equivalence transformations, and construct the tangent space corresponding to transformations that do not leave the orbit of $\mathcal{K}$-equivalence.
Consider a one-parameter family of transformations given by $A_t(\vb{k})$ and $\vb*{\phi}_t(\vb{k})$ where $t \in \mathbb{R}$, and $A_0(\vb{k}) = \mathbbm{1}$ and $\vb*{\phi}_0(\vb{k}) = \vb{k}$ corresponding to the identity transformation.
Note that by definition $\vb*{\phi}_t$ fixes the origin, i.e., $\vb*{\phi}_t(\vb{0})=\vb{0}$ holds for every $t$.
An infinitesimal transformation on $\vb{d}(\vb{k})$ reads
\begin{equation}
\label{eqn:Ktrf}
    \left.\frac{d}{dt}\vb{d}_t(\vb{k})\right|_{t=0} =
    \left.\frac{d}{dt}\left[A_t(\vb{k})\vb{d}_0\!\left(\vb*{\phi}_t\left(\vb{k}\right)\right)\right]\right|_{t=0} = B(\vb{k}) \vb{d}_0(\vb{k}) + J(\vb{k}) \vb{c}(\vb{k}),
\end{equation}
where
\begin{eqnarray}
    B(\vb{k}) &=& \left.\frac{d}{dt}A_t(\vb{k})\right|_{t=0},\nonumber\\
    \vb{c}(\vb{k}) &=& \left.\frac{d}{dt}\vb*{\phi}_t(\vb{k})\right|_{t=0}, \\
    J(\vb{k})_{ij} &=& \frac{\partial d_{0i}}{\partial k_j}(\vb{k}).\nonumber
\end{eqnarray}
The functions $B(\vb{k})$ and $\vb{c}(\vb{k})$ are arbitrary matrix and vector-valued smooth functions (power series) with $\vb{c}(\vb{0})=\vb{0}$ and $J(\vb{k})$ is the Jacobian of $\vb d(\vb k)$. All the infinitesimal transformations that can be written in the form of \eqref{eqn:Ktrf} remain in first order in the orbit of $\mathcal{K}$-equivalence, forming the $\mathcal{K}$-tangent space.
Power series that cannot be represented as \eqref{eqn:Ktrf} form \emph{transverse perturbations} of $\vb{d}_0$, which leave the orbit of $\mathcal{K}$-equivalence.
The space of transverse perturbations fixing the origin is formed by the quotient of the space of all power series with zero constant term in the space of power series of the form \eqref{eqn:Ktrf}. The dimension of the quotient space is called the $\mathcal{K}$-codimension of the germ $\vb{d}(\vb{k})$.

However, we want to consider perturbations which possibly move the origin.
For it we need the extended tangent space ($\mathcal{K}_e$-tangent space), defined in the same way as the right side of \eqref{eqn:Ktrf}, but without the restriction $\vb{c}(\vb{0})=\vb{0}$.
Hence we allow for perturbed functions with $\vb{d}_{\vb{t}}(\vb{0}) \neq \vb{0}$ for $\vb{t}\neq \vb{0}$.
The space of all  perturbations transverse to the extended tangent space is formed by the quotient of the space of all power series (with possibly non-zero constant term) with the space of power series of the form \eqref{eqn:Ktrf}.
The dimension of the quotient space is called the $\mathcal{K}_e$-codimension of the germ $\vb{d}(\vb{k})$, the subscript refers to `extended'.

In practice this calculation can be carried out in the following way:
First we show that all vector valued monomials above a certain degree can be generated by a suitable choice of $B(\vb{k})$ and $\vb{c}(\vb{k})$.
In the next step, we restrict $B(\vb{k})$ and $\vb{c}(\vb{k})$ to be polynomials of this maximal order, restricting them to a finite dimensional subspace of the vector and matrix valued polynomial ring.
Then we evaluate \eqref{eqn:Ktrf} for a set of basis functions in this space, and expand the result (also restricted to this finite dimensional subspace) in a basis, resulting in a matrix equation between the free coefficients in $B(\vb{k})$ and $\vb{c}(\vb{k})$ and the coefficients of the resulting $\frac{d}{dt}\vb{d}_t(\vb{k})$.
Calculating the rank deficiency of this matrix (the dimension of its null space) gives the codimension, and a basis of the complement of its image yields the transverse perturbations.

\subsection{Example: double-Weyl node}

For the following example we compute the $\mathcal{K}_e$-tangent space, we give a basis for the transverse perturbations, and conclude that the $\mathcal{K}_e$-codimension is 4 without requiring any symmetry:
\begin{equation}
\label{eqn:double_Weyl_nf}
    \vb{d}_0\!\left(\vb{k}\right)= 
    \begin{pmatrix}
        k_x^2-k_y^2\\
        k_x k_y\\
        k_z\\
    \end{pmatrix}.
\end{equation}

For the calculation we explicitly apply Eq.~\eqref{eqn:Ktrf}, and enumerate the vector valued power series that can be written in the form of the right hand side, and consequently, are in the $\mathcal{K}_e$ tangent space.
Power series that cannot be written this way form the transverse perturbations.
As a final step, we find a basis of the transverse perturbations by eliminating perturbations that can be written as a linear combination of transverse perturbations and elements of the tangent space.

The first term on the right hand side of Eq.~\eqref{eqn:Ktrf} yields expressions in all components ($i=1,2,3$) of $\left.\frac{d}{dt}\vb{d}_t(\vb{k})\right|_{t=0}$ such as
\begin{equation}
\label{eqn:B_ideal}
    \left[B(\vb{k}) \vb{d}_0(\vb{k})\right]_i = B_{i1}(k_x,k_y,k_z)(k_x^2-k_y^2)+B_{i2}(k_x,k_y,k_z)k_xk_y+B_{i3}(k_x,k_y,k_z)k_z,
\end{equation}
where $B_{ij}$ are arbitrary functions of $(k_x,k_y,k_z)$.
Scalar functions of this form are the elements of the \emph{ideal} generated by the components of $\vb d(\vb k)$ (for a brief introduction see the appendix of \cite{Pinter2022}).
Vector valued functions with components of this form span part of the $\mathcal{K}_e$ tangent space.
In this specific case, the last term $B_{i3}(k_x,k_y,k_z)k_z$ ensures that every monomial in every component with a $k_z$ factor is in the ideal.
Moreover, the second term $B_{i2}(k_x,k_y,k_z)k_x k_y$ gives every monomial with a $k_x k_y$ factor.
Therefore, only terms that are the pure powers of $k_x$ and $k_y$ are left as possible transverse perturbations.
All pure $k_x$ or $k_y$ powers above second order are in the ideal, as for $a \geq 0$ they can be written in the form of \eqref{eqn:B_ideal}:
\begin{eqnarray}
    k_x^{a+3}&=&k_x^{a+1}\left(k_x^2-k_y^2\right)+k_x^ak_y\left(k_xk_y\right),\\
    k_y^{a+3}&=&-k_y^{a+1}\left(k_x^2-k_y^2\right)+k_xk_y^a\left(k_xk_y\right).
\end{eqnarray}
The remaining monomials that are possible components of transverse perturbations are $1$, $k_x$, $k_y$, $k_x^2$ and $k_y^2$. The last two are not linearly independent, as their difference $k_x^2 - k_y^2$ is a tangent perturbation, which can be written in the form \eqref{eqn:B_ideal}, therefore, we can remove $k_y^2$ from the list of basis functions. Up to this point, a potential basis of transverse perturbations is
\begin{equation}
    \left\langle \begin{pmatrix}
1\\
0\\
0
\end{pmatrix}, \begin{pmatrix}
k_x\\
0\\
0
\end{pmatrix},
\begin{pmatrix}
k_y\\
0\\
0
\end{pmatrix},
\begin{pmatrix}
k_x^2\\
0\\
0
\end{pmatrix},
\begin{pmatrix}
0\\
1\\
0
\end{pmatrix}, \begin{pmatrix}
0\\
k_x\\
0
\end{pmatrix},
\begin{pmatrix}
0\\
k_y\\
0
\end{pmatrix},
\begin{pmatrix}
0\\
k_x^2\\
0
\end{pmatrix},
\begin{pmatrix}
0\\
0\\
1
\end{pmatrix}, \begin{pmatrix}
0\\
0\\
k_x
\end{pmatrix},
\begin{pmatrix}
0\\
0\\
k_y
\end{pmatrix},
\begin{pmatrix}
0\\
0\\
k_x^2
\end{pmatrix}\right\rangle,
\end{equation}
since we showed that all other functions are a linear combination of these and functions in the $\mathcal{K}_e$-tangent space.

Next, we move on to the second term in Eq.~\eqref{eqn:Ktrf}. The possible expressions have the form
\begin{equation}
\label{eqn:charge_2_second_term}
J(\vb{k}) \vb{c}(\vb{k}) =  
c_1(k_x,k_y,k_z)
\begin{pmatrix}
2k_x\\
k_y\\
0
\end{pmatrix}+
c_2(k_x,k_y,k_z)
\begin{pmatrix}
-2k_y\\
k_x\\
0
\end{pmatrix}+
c_3(k_x,k_y,k_z)
\begin{pmatrix}
0\\
0\\
1
\end{pmatrix},
\end{equation}
where the vectors are the columns of the Jacobian $J(\vb k)$ and $c_i$ are arbitrary functions.
The third term gives any monomial in the third component, therefore we only need to consider perturbations that are zero in the third component as possible basis vectors for transverse perturbations.

The two vectors below with second order monomials in their components are in the $\mathcal{K}_e$-tangent space, as they can be written in the form \eqref{eqn:Ktrf} using terms from both \eqref{eqn:B_ideal} and \eqref{eqn:charge_2_second_term}:
\begin{eqnarray}
\begin{pmatrix}
k_x^2\\
0\\
0
\end{pmatrix}&=&
    \frac{k_x}{2}
    \begin{pmatrix}
2k_x\\
k_y\\
0
\end{pmatrix}-
\frac{1}{2}
\begin{pmatrix}
0\\
k_xk_y\\
0
\end{pmatrix}\\
\begin{pmatrix}
0\\
k_x^2\\
0
\end{pmatrix}&=&
    k_x
    \begin{pmatrix}
-2k_y\\
k_x\\
0
\end{pmatrix}+2
\begin{pmatrix}
k_xk_y\\
0\\
0
\end{pmatrix}.
\end{eqnarray}
At this point this leaves the potential basis of $\mathcal{K}_e$-transverse perturbations:
\begin{equation}
    \left\langle \begin{pmatrix}
1\\
0\\
0
\end{pmatrix}, \begin{pmatrix}
k_x\\
0\\
0
\end{pmatrix},
\begin{pmatrix}
k_y\\
0\\
0
\end{pmatrix},
\begin{pmatrix}
0\\
1\\
0
\end{pmatrix}, \begin{pmatrix}
0\\
k_x\\
0
\end{pmatrix},
\begin{pmatrix}
0\\
k_y\\
0
\end{pmatrix}
\right\rangle.
\end{equation}
The two vectors with $k_y$ can be expressed using the two vectors with $k_x$ and the first two columns of the Jacobian $J(\vb k)$ in \eqref{eqn:charge_2_second_term} which are in the $\mathcal{K}_e$-tangent space:
\begin{eqnarray}
\begin{pmatrix}
k_y\\
0\\
0
\end{pmatrix}&=&
    -\frac{1}{2}
    \begin{pmatrix}
-2k_y\\
k_x\\
0
\end{pmatrix}+\frac{1}{2}
\begin{pmatrix}
0\\
k_x\\
0
\end{pmatrix},\\
\begin{pmatrix}
0\\
k_y\\
0
\end{pmatrix}&=&
    \begin{pmatrix}
2k_x\\
k_y\\
0
\end{pmatrix}
    -2
\begin{pmatrix}
k_x\\
0\\
0
\end{pmatrix}.
\end{eqnarray}
Thus, only 4 vectors give independent $\mathcal{K}_e$-transverse perturbations.

A possible basis of $\mathcal{K}_e$-transverse perturbations is given by
\begin{equation}\label{eq:transvbasis}
    \left\langle \begin{pmatrix}
1\\
0\\
0
\end{pmatrix}, \begin{pmatrix}
0\\
1\\
0
\end{pmatrix},
\begin{pmatrix}
k_x\\
0\\
0
\end{pmatrix},
\begin{pmatrix}
0\\
k_x\\
0
\end{pmatrix}\right\rangle.
\end{equation}
As we showed above, any power series can be constructed as a linear combination of these basis functions with a tangent function of the form \eqref{eqn:Ktrf}.
It is easy to see that no smaller set satisfies this condition, because no linear combination of these basis functions is a tangent function of the form \eqref{eqn:Ktrf}.

We point out that the basis vectors of $\mathcal{K}_e$-transverse perturbations are not unique; they can be shifted with tangent perturbations. A different basis reads
\begin{equation}
    \left\langle \begin{pmatrix}
1\\
0\\
0
\end{pmatrix}, \begin{pmatrix}
0\\
1\\
0
\end{pmatrix},
\begin{pmatrix}
k_x\\
0\\
0
\end{pmatrix},
\begin{pmatrix}
k_y\\
0\\
0
\end{pmatrix}\right\rangle.
\end{equation}

We note that there are other inequivalent double-Weyl nodes in the more general sense, that is, degeneracy points of charge $\pm 2$, e.g.
\begin{equation}
    \label{eqn:higher_order_dw}
    \vb{d}_n\!\left(\vb{k}\right)=
    \begin{pmatrix}
    k_x^{2n}-k_y^{2n}\\
    k_x k_y\\
    k_z
    \end{pmatrix}
\end{equation}
for $n > 1$ integers, but these have higher codimension, therefore, require more fine-tuning.

\subsection{Unfoldings and versality}

Perturbations or deformations of germs are formulated via the notion of \emph{unfoldings}. An unfolding of $\vb{d}(\vb{k})$ is a germ in form $\vb{D}(\vb{k}, \vb{t})=(\vb{d}(\vb{k}, \vb{t}), \vb{t})$, where $\vb{t}\in \mathbb{R}^n$ for some $n$ and $\vb{d}(\vb{k}, \vb{t}=\vb{0}) = \vb{d}(\vb{k})$. Its first component $\vb{d}(\vb{k}, \vb{t})$ is the perturbation or deformation corresponding to the control parameter $\vb{t}$.
Note that for a fixed $\vb{t}$ value, the perturbation $\vb{k} \mapsto \vb{d}(\vb{k}, \vb{t})$ is a map of $\vb{k}$, but not a germ. However, $\vb{D}(\vb{k}, \vb{t})$ is a germ thanks to the inclusion of the trivial $\vb{t}$ component. We fix a sufficiently small representative map of the unfolding $\vb{D}(\vb{k}, \vb{t})$ with $|\vb{k}| < \epsilon$ and $|\vb{t}| < \delta$, such that $\vb{d}(\vb{k}, \vb{t})=0$ has no solutions with $|\vb{t}| <\delta $ and $|\vb{k}| = \epsilon$, i.e., the gap does not close on the boundary of the $\epsilon$-ball in the $\vb{k}$-space. This makes the analysis shown by Fig. 2. in the main text precise.

An unfolding $\vb{D}_v\!\left(\vb{k}, \vb{t}\right)=(\vb{d}_v(\vb{k}, \vb{t}), \vb{t})$ ($\vb{t} \in \mathbb{R}^m$) is \emph{versal} (more precisely, $\mathcal{K}$-versal), if every other unfolding can be induced from it up to contact equivalence: given any unfolding $\vb{D}(\vb{k}, \vb{t})$ ($\vb{t} \in \mathbb{R}^n$), there exists a base-change function (pull-back) $\vb{h} : \mathbb{R}^n \to \mathbb{R}^m$ such that
\begin{equation}
    \widetilde{\vb{D}}(\vb{k}, \vb{t}) := (\vb{d}_v\!\left(\vb{k}, \vb{h}(\vb{t})), \vb{t}\right)
\end{equation}
agrees with $\vb{D}(\vb{k}, \vb{t})$ up to \emph{contact equivalence of unfoldings}, see Ref.~\onlinecite{Wall,Damon} for precise definition. It is also \emph{miniversal}, if there exists no versal unfolding with a lower number of parameters.

By the \emph{unfolding theorem}~\cite{Wall, Damon}, a miniversal unfolding can be constructed as follows. We choose a basis for a perturbations transverse to the $\mathcal{K}_e$-tangent space, the number of basis elements is the $\mathcal{K}_e$-codimension of $\vb{d}(\vb{k})$ by definition. Adding the basis elements to $\vb{d}(\vb{k})$  with arbitrary parameters, we get an unfolding. The unfolding theorem states that this unfolding is miniversal.
 
In our example, using the basis given in Equation~\eqref{eq:transvbasis} (and $\vb{t}$ as a vector of parameters), we get the unfolding
\begin{equation}
\label{eqn:double_Weyl_unfolding}
    \vb{d}_v\!\left(\vb{k}, \vb{t}\right)= 
    \begin{pmatrix}
        k_x^2-k_y^2\\
        k_x k_y\\
        k_z\\
    \end{pmatrix}
    + t_1 \begin{pmatrix}
    1\\
    0\\
    0
    \end{pmatrix}
    + t_2 \begin{pmatrix}
    0\\
    1\\
    0
    \end{pmatrix}
    + t_3 \begin{pmatrix}
    k_x\\
    0\\
    0
    \end{pmatrix}
    + t_4 \begin{pmatrix}
    0\\
    k_x\\
    0
    \end{pmatrix},
\end{equation}
which is miniversal by the unfolding theorem. 
In other words, any double-Weyl node (of the most generic kind) and any of its perturbations can be written in the form \eqref{eqn:double_Weyl_unfolding} after an appropriate reparametrization of momentum space, the space of $2\times 2$ hermitian matrices, and parameter space, in the sense of contact equivalence.
We note that although the construction in Section~\ref{ss:conteq} involving the $\mathcal{K}_e$-tangent space suggests \emph{infinitesimal versality}, the statement that this unfolding is indeed versal  is a nontrivial theorem (unfolding theorem). 
We illustrate the versality on two examples, i.e. we show for some specific perturbations, how they can be induced from the unfolding  given by Equation~\eqref{eqn:double_Weyl_unfolding}. 

\subsubsection*{Example 1 of transformation to the versal unfolding}
First, we demonstrate on a one-parameter example that the perturbations of the component $d_z$ and perturbations containing $k_z$ do not yield qualitatively new perturbations.
We consider the following specific example:
\begin{equation}\label{eq:unfold_eqv_d0}
    \vb d_0(\vb k,t)=
    \begin{pmatrix}
        k^2_x -k^2_y\\
        k_xk_y\\
        k_z
    \end{pmatrix}+t
    \begin{pmatrix}
        k_yk_z\\
        1\\
        k^2_x
    \end{pmatrix}.
\end{equation}
This form can be transformed into the versal unfolding by a series of simple steps. The transformation of the domain ($\vb{k}$-space) reads
\begin{equation}\label{eq:unfold_eqv_phi}
\vb*{\phi}_t(\vb k)=
    \begin{pmatrix}
        k_x\\
    k_y\\
    k_z-tk^2_x
    \end{pmatrix}.
\end{equation}
It depends analytically on $t$ and it is the identity for $t=0$.
It simplifies the third component of the unfolding
\begin{equation}\label{eq:unfold_eqv_d1}
    \vb{d}_1(\vb k,t)=\vb d_0(\vb*{\phi}_t(\vb k),t)=
    \begin{pmatrix}
        k_x^2-k_y^2+tk_y(k_z-tk_x^2)\\
        k_xk_y+t\\
        k_z
    \end{pmatrix}.
\end{equation}
The transformation of the image ($\vb d$-space) reads
\begin{eqnarray}\label{eq:unfold_eqv_psi1}
\vb*\psi_{1,t}(\vb k, \vb d)=
    \begin{pmatrix}
        d_x-tk_yd_z\\
        d_y\\
        d_z
    \end{pmatrix}.
\end{eqnarray}
It depends on $\vb k$ and t analytically and it is the identity for $t=0$.
This is a linear transformation of $\vb{d}$-space, and could be expressed using our earlier notation as $\vb*\psi_{1,t}(\vb k, \vb d) = A_t (\vb{k}) \vb{d}$ with a parameter-dependent matrix
\begin{equation}
    A_t (\vb{k})=\begin{pmatrix}
        1&0&-tk_y\\
        0&1&0\\
        0&0&1
    \end{pmatrix}.
\end{equation}
This transformation cancels one term in the first component of the unfolding
\begin{equation}\label{eq:unfold_eqv_d2}
    \vb d_2(\vb k,t)=\vb*\psi_{1,t}(\vb k,\vb{d}_1(\vb k,t))=
    \begin{pmatrix}
        k_x^2-k_y^2-t^2k_x^2k_y\\
        k_xk_y+t\\
        k_z
    \end{pmatrix}.
\end{equation}
A second transformation in the image reads
\begin{equation}\label{eq:unfold_eqv_psi2}
    \vb*\psi_{2,t}(\vb k, \vb d)=
    \begin{pmatrix}
        d_x+t^2k_xd_y\\
        d_y\\
        d_z
    \end{pmatrix},
\end{equation}
which further simplifies the unfolding:
\begin{equation}\label{eq:unfoldingequivalence}
   \vb d_3(\vb k,t)=\vb*\psi_{2,t}(\vb k,\vb d_2(\vb k,t))=
    \begin{pmatrix}
        k_x^2-k_y^2\\
        k_xk_y\\
        k_z
    \end{pmatrix}+t
    \begin{pmatrix}
        0\\
        1\\
        0
    \end{pmatrix}+t^3
    \begin{pmatrix}
        k_x\\
        0\\
        0
    \end{pmatrix}.
\end{equation}
So far we only performed contact-equivalence transformations of unfoldings, hence the unfolding $\vb{d}_3$ is equivalent to the $\vb{d}_0$ we started with.
Now comparing Eq.~\eqref{eqn:double_Weyl_unfolding} and Eq.~\eqref{eq:unfoldingequivalence}  yields
\begin{equation}\label{eq:unfold_eqv_d3}
    \vb d_3(\vb k, t)=\vb d_v(\vb{k},\vb{h}(t))
\end{equation}
with the pull-back $\vb{h}(t)$ to the four dimensional parameter space of the versal unfolding:
\begin{equation}\label{eq:unfold_eqv_pullback}
\vb{h}(t) = 
    \begin{pmatrix}
        0\\
        t\\
        t^3\\
        0
    \end{pmatrix}.
\end{equation}

\subsubsection*{Example 2 of transformation to the versal unfolding}
Now we show a second example, where we include all six perturbations in the first two components up to linear order in $k_x$ and $k_y$:
\begin{equation}
    \vb{d}_0\!\left(\vb{k}, \vb{t}\right)= 
    \begin{pmatrix}
        k_x^2-k_y^2\\
        k_x k_y\\
        k_z\\
    \end{pmatrix}
    + t_1 \begin{pmatrix}
    1\\
    0\\
    0
    \end{pmatrix}
    + t_2 \begin{pmatrix}
    0\\
    1\\
    0
    \end{pmatrix}
    + t_3 \begin{pmatrix}
    k_x\\
    0\\
    0
    \end{pmatrix}
    + t_4 \begin{pmatrix}
    0\\
    k_x\\
    0
    \end{pmatrix}
    + t_5 \begin{pmatrix}
    k_y\\
    0\\
    0
    \end{pmatrix}
    + t_6 \begin{pmatrix}
    0\\
    k_y\\
    0
    \end{pmatrix}.
\end{equation}
To get rid of the $k_y$ terms, we shift the domain by
\begin{equation}
    \vb*{\phi}_{\vb{t}}(\vb{k})=\begin{pmatrix}
        k_x-t_6\\
        k_y+\frac{t_5}{2}\\
        k_z
    \end{pmatrix},
\end{equation}
which results in
\begin{align}
\nonumber
    \vb{d}_1(\vb k, \vb{t})=\vb d_0(\vb*{\phi}_{\vb{t}}(\vb k),\vb{t})=&
        \begin{pmatrix}
        k_x^2-k_y^2\\
        k_x k_y\\
        k_z\\
    \end{pmatrix}
    + \left(t_1-t_3t_6+t_6^2+\frac{t_5^2}{4}\right) \begin{pmatrix}
    1\\
    0\\
    0
    \end{pmatrix}
    + \left(t_2-t_4 t_6\right) \begin{pmatrix}
    0\\
    1\\
    0
    \end{pmatrix}\\
    &+(t_3-2t_6) \begin{pmatrix}
    k_x\\
    0\\
    0
    \end{pmatrix}
    + \left(t_4+\frac{t_5}{2}\right) \begin{pmatrix}
    0\\
    k_x\\
    0
    \end{pmatrix}.
\end{align}
From this form we can read off the pull-back
\begin{equation}
\vb h(\vb t)=\begin{pmatrix}
    t_1-t_3t_6+t_6^2+\frac{t_5^2}{4}\\
    t_2-t_4 t_6\\
    t_3-2t_6\\
    t_4+\frac{t_5}{2}
\end{pmatrix},    
\end{equation}
to satisfy the identity $\vb{d}_1(\vb k, \vb t)=\vb d_v(\vb{k},\vb{h}(\vb t))$. This calculation demonstrates that the remaining linear terms do not add qualitatively new perturbations to the normal form.

\subsection{Relation to codimension of bifurcation sets}

The abstract notion of $\mathcal{K}_e$-codimension of a germ $\vb{d}(\vb{k})$ can be realised as the codimension of a submanifold in the space  $\mathbb{R}^n$ of control parameters $\vb{t}$ of the miniversal unfolding $\vb{D}(\vb{k}, \vb{t})$.

For a generic fixed control parameter $\vb{t}$ the perturbed map $\vb{k} \mapsto \vb{d}(\vb{k}, \vb{t})$  has only ordinary Weyl points, i.e. if $\vb{d}(\vb{k}, \vb{t})=\vb{0}$ for a $\vb{k}$, then the Jacobian of $\vb{k} \mapsto \vb{d}(\vb{k}, \vb{t})$ has  maximal rank 3.
The exceptional control parameters $\vb{t}$ for which this is not true form the $\mathcal{K}$-\emph{bifurcation set} in $\mathbb{R}^n$. Its strata correspond to the non-generic degeneracy patterns which can appear in any perturbation of $\vb{d}$, see Appendix of \cite{Pinter2022}. The bifurcation set 
separates $\mathbb{R}^n$ into regions with possibly different number of ordinary Weyl points, and the merging processes correspond to the motions of $\vb{t}$ along curves intersecting the $\mathcal{K}$-bifurcation set.

The $\mathcal{K}_e$-codimension of each appearing degeneracy point is equal to the codimension of the corresponding stratum in the control parameter space  $\mathbb{R}^n$ (see [Mond, Thm. 5.6. and Prop. 5.1.], formulated for left-right equivalence).  In our example $\vb{t} \in \mathbb{R}^4$. The bifurcation set contains the origin (codim 4) corresponding to the double Weyl point ($\mathcal{K}_e$-codim 4). The bifurcation set also contains the $\vb{t}$ parameter values corresponding to three node processes ($\mathcal{K}_e$-codim 2), and they form surfaces in $\mathbb{R}$ (codimension 2). Finally, the control parameter values corresponding to pairwise annihilation processes ($\mathcal{K}_e$-codim 1) form a 3 dimensional submanifold in $\mathbb{R}^4$ (codimension 1). 

Although without symmetry the control parameter space is 4 dimensional, its dimension decreases if we introduce symmetry, e.g. $C_2 \mathcal{T}$.  On Figs. \ref{fig:phasediagram_SrSi} and \ref{fig:phase_graphene} the bifurcation set contains the green lines (three-node processes) and the magenta lines (pairwise annihilation), both having codimension 1 in the 2 dimensional control parameter space, according to the fact that the codimension of both process is 1 with $C_2 \mathcal{T}$ symmetry, cf. Table~\ref{tab:propdegpoints}. The origin, also contained by the bifurcation set, has codimension 2 in the plane, equals to the codimension of the double Weyl point with $C_2 \mathcal{T}$ symmetry. Fig.~\ref{fig:Weylbreaking} shows the bifurcation set of a versal, but not miniversal unfolding with 3 control parameters in the presence of $C_2 \mathcal{T}$ symmetry. We can see that every stratum of the bifurcation set has 1 bigger dimension than on the 2 dimensional pictures, hence the codimensions remain the same. 

\subsection{Relation to Ref.~\texorpdfstring{\onlinecite{Teramoto2017}}{2}}
We note that Ref.~\onlinecite{Teramoto2017} uses a different equivalence relation, where two Hamiltonians are considered equivalent if there exist smooth $\vb*{\phi}(\vb{k}): \mathbb{R}^3\mapsto\mathbb{R}^3$ and $U(\vb{k}): \mathbb{R}^3\mapsto\textnormal{SU}(2)$ such that
\begin{equation}
    H'(\vb{k}) = U(\vb{k}) H(\vb*{\phi}(\vb{k})) U(\vb{k})^{-1}.
\end{equation}
In our formalism, this is equivalent to the restriction of $A \in \textnormal{SO}(3)$.
This is a stronger equivalence relation, resulting in continuum many classes in certain situations.
We justify our choice of $\mathcal{K}$-equivalence in the space of the $\vb{d}$-vectors, because it allows for the most generic deformations of two-band Hamiltonians without introducing additional gap closings, while the stronger equivalence relation is sensitive to the detailed energetics of the band structure near the degeneracy point.
For example, in the classification of Ref.~\onlinecite{Teramoto2017} the double Weyl point splits into continuum many classes.

\section{Symmetry constraints}

We modify the above construction of the tangent space, transverse perturbations, and unfoldings to be applicable in the presence of symmetries, by restricting to the subspace of functions that respect the prescribed symmetries.

A general symmetry poses a restricion on $\vb{d}(\vb{k})$ of the form
\begin{equation}
    V \vb{d}(\vb{k}) = \vb{d}(R \vb{k}),
\end{equation}
where $V$ and $R$ are $3\times 3$ orthogonal matrices acting in $\vb{d}$ and $\vb{k}$ space respectively.
We demand that the contact equivalence transformation of Eq.~\eqref{eqn:KLeq_trf} preserves this symmetry, which results in the constraints on the symmetry-compatible transformations described by $A(\vb{k})$ and $\vb*{\phi}\!\left(\vb{k}\right)$:
\begin{align}
    V A(\vb{k}) =& A(R\vb{k}) V, \\
    R\, \vb*{\phi}\!\left(\vb{k}\right) =& \vb*{\phi}\!\left(R \vb{k}\right).
\end{align} 
Constraints of the same form also apply to $B(\vb{k})$ and $\vb{c}(\vb{k})$.
The computation of the codimension and normal form follows the same procedure as without symmetry, except that the spaces of available functions are restricted.

Although in our examples these restrictions are straightforward (see below), for the general theory we refer to Ref.~\onlinecite{Roberts}, ~\onlinecite{WallJet}.
Note that this restriction guarantees that \emph{any} symmetric $\vb{d}(\vb{k})$ is mapped to a symmetric one by the allowed contact equivalence transformations.
There are cases where two symmetric $\vb{d}(\vb{k})$'s are connected by an unrestricted contact equivalence transformation, but not by a restricted one.
We give a non-trivial example of this situation in \eqref{eqn:rotated_double_weyl}.
%

\begin{table}[ht]
    \setlength{\extrarowheight}{3pt}
	\begin{tabularx}{\textwidth}{|>{\raggedright\arraybackslash}X|>{\raggedright\arraybackslash}X|>{\raggedright\arraybackslash}X|>{\raggedright\arraybackslash}X|>{\raggedright\arraybackslash}X|}
	    \hline
		&(ordinary) Weyl point & Quadratic Weyl point & Cubic Weyl point & Double Weyl point \\
  \hline\hline
   Normal form without symmetry & 
   $\begin{pmatrix}
      k_x\\
      k_y\\
      k_z
  \end{pmatrix}$ & 
  $\begin{pmatrix}
      k_x^2+t\\
      k_y\\
      k_z
  \end{pmatrix}$& 
  $\begin{pmatrix}
      k_x\\
      k_y^3+t_1+t_2k_y\\
      k_z
  \end{pmatrix}$& 
  $\begin{pmatrix}
      k_x^2-k_y^2+t_1+t_2k_x\\
      k_xk_y+t_3+t_4k_x\\
      k_z
  \end{pmatrix}$\\
  \hline
  Codimension without symmetry
  & 0& 1& 2& 4\\
  \hline
  Normal form with $C_{2y} \mathcal{T}$ symmetry & $\begin{pmatrix}
      k_x\\
      k_y\\
      k_z
  \end{pmatrix}$ & 
  $\begin{pmatrix}
      k_x^2+t\\
      k_y\\
      k_z
  \end{pmatrix}$& 
  $\begin{pmatrix}
      k_x\\
      k_y^3+tk_y\\
      k_z
  \end{pmatrix}$& 
  $\begin{pmatrix}
      k_x^2-k_y^2+t_1+t_2k_x\\
      k_xk_y\\
      k_z
  \end{pmatrix}$\\
  \hline
  Codimension with $C_{2y} \mathcal{T}$ symmetry
  & 0& 1& 1& 2\\
  \hline
  Charge (Chern number, local degree)
   & $\pm 1$& 0& $\pm 1$& $\pm 2$\\
  \hline
  Birth quota (local multiplicity)~\cite{Teramoto2017,Pinter2022}
   & 1& 2& 3& 4\\
  \hline
  Dispersion
   & {linear in every direction}& {quadratic in one direction}& {cubic in one direction}& {quadratic in a plane}\\
  \hline
  Remark& 
   {Generic two-fold degeneracy point without symmetry.} & {Generic Weyl-point process without symmetry, pairwise creation and annihilation} & {3-point process, which is stabilized by the $C_2T$ symmetry.} & {This is the most generic doubly charged point} \\
  \hline
  \end{tabularx}
	\caption{Some properties of the simplest isolated two-fold degeneracy points. The degeneracies are identified with contact equivalence classes of germs. Here, these representatives are  formulated as germs from $\mathbb{R}^3$  to $\mathbb{R}^3$, but only the non-linear part is essential. In this sense the quadratic, respectively, the cubic Weyl point is essentially determined by the $\mathbb{R} \to \mathbb{R}$ germs $f(x)=x^2$, respectively, $g(x)=x^3$. In the same sense, the double Weyl point is essentially an $\mathbb{R}^2 \to \mathbb{R}^2$ germ, as it appears in the case of the bilayer graphene (called Dirac point in two dimension). \label{tab:propdegpoints}}
\end{table}

In the specific case of of $C^y_2\mathcal{T}$ symmetry introduced in the main text, the function $\vb{d}(\vb{k})$ is restricted as:
\bean\label{eq:Hsymm}
d_x(k_x,-k_y,k_z)&=&d_x(k_x,k_y,k_z)\nonumber\\
d_y(k_x,-k_y,k_z)&=&-d_y(k_x,k_y,k_z)\\
d_z(k_x,-k_y,k_z)&=&d_z(k_x,k_y,k_z)\nonumber.
\eean
In the notation above this corresponds to $V = R = \operatorname{diag}(1, -1, 1)$.
We summarize the results of codimension calculations with or without $C_2\mathcal{T}$ symmetry in Table~\ref{tab:propdegpoints}.

\subsection{Double-Weyl nodes with symmetry}
\subsubsection{With \texorpdfstring{$C^y_2\mathcal{T}$}{C2T} symmetry}
We find the miniversal unfolding of \eqref{eqn:double_Weyl_nf} (which already satisfies the symmetry constraint) restricted by $C_2\mathcal{T}$ symmetry to be (Eq.~(1) of the main text)
\begin{equation}
\label{eqn:double_Weyl_sym_unfolding}
    \vb{d}_v\!\left(\vb{k}, \vb{t}\right)= 
    \begin{pmatrix}
        k_x^2-k_y^2\\
        k_x k_y\\
        k_z\\
    \end{pmatrix}
    + t_1 \begin{pmatrix}
    1\\
    0\\
    0
    \end{pmatrix}
    + t_2 \begin{pmatrix}
    k_x\\
    0\\
    0
    \end{pmatrix}.
\end{equation}
This is the restricted version of the versal unfolding described in Eq.~\eqref{eqn:double_Weyl_unfolding} with the $C^y_2\mathcal{T}$ symmetry.
This shows that with the $C^y_2\mathcal{T}$ symmetry the double-Weyl node has codimension 2.

\subsubsection{Example of transformation to the normal form with \texorpdfstring{$C^y_2\mathcal{T}$}{C2T} symmetry}
To demonstrate the versality with the $C^y_2\mathcal{T}$ symmetry, we show on an example that other perturbations that are naively independent, can be brought to this form.
Consider
\begin{equation}
\label{eq:doubleWP}
\vec d_0(\vb{k}, \vb{t})=
\begin{pmatrix}
k_x^2-k_y^2\\
k_x k_y\\
k_z
\end{pmatrix}+t_1
\begin{pmatrix}
1 \\
0\\
0
\end{pmatrix}+t_2
\begin{pmatrix}
k_x \\
0\\
0
\end{pmatrix}+t_3
\begin{pmatrix}
0\\
k_y\\
0
\end{pmatrix},
\end{equation}
where we consider every symmetry-preserving monomial up to linear order (except $k_z$) in the first two components.
This perturbation can be reduced with the shift of the $k_x$ coordinate 
\begin{equation}   
\label{eq:xshift}
\vb*{\phi}_{\vb{t}}(\vb{k})= \begin{pmatrix}
    k_x-t_3\\
    k_y\\
    k_z
\end{pmatrix}.
\end{equation}
This shift preserves the $C^y_2\mathcal{T}$ for every $t$. The resulting reduced 2-parameter perturbation reads
\begin{equation}
\label{eq:doubleWP2}
\vb{d}_1( \vb{k}, \vb{t}) = \vec d_0( \vb*{\phi}_{ \vb{t} }(\vb{k}), \vb{t})=
\begin{pmatrix}
k_x^2-k_y^2\\
k_x k_y\\
k_z
\end{pmatrix}+\left(t_1-t_2t_3 + t^2_3\right)
\begin{pmatrix}
    1\\
    0\\
    0
\end{pmatrix}+(t_2- 2 t_3)
\begin{pmatrix}
    k_x\\
    0\\
    0
\end{pmatrix}.
\end{equation}
Eq.~\eqref{eq:doubleWP2} implicitly defines the pull-back $\vb h(\vb{t})$ for which $\vb{d}_1(\vb k,\vb t)=\vb d_v(\vb k,\vb h(\vb t))$ is satisfied:
\begin{eqnarray}\label{eq:pullback}
    \vb h (\vb t) =
    \begin{pmatrix}
t_1 - t_2t_3 + t_3^2\\
t_2 - 2 t_3
\end{pmatrix}.
\end{eqnarray}

\subsubsection{With \texorpdfstring{$C^y_2\mathcal{T}$ and $C_{2z}$}{C2yT and C2z} symmetry}
If the symmetry is enhanced by adding $C_{2z}$ symmetry, represented trivially on the internal degrees of freedom, we get the additional symmetry constraints
\bean
d_x(-k_x,-k_y,k_z)&=&d_x(k_x,k_y,k_z)\nonumber\\
d_y(-k_x,-k_y,k_z)&=&d_y(k_x,k_y,k_z)\\
d_z(-k_x,-k_y,k_z)&=&d_z(k_x,k_y,k_z).\nonumber
\eean
The resulting normal form is
\bean
\vec d_v(\vb{k}, t)=
\begin{pmatrix}
k_x^2-k_y^2\\
k_x k_y\\
k_z
\end{pmatrix}+t
\begin{pmatrix}
1 \\
0\\
0
\end{pmatrix},
\eean
showing that the symmetry makes the double-Weyl point codimension 1, and the two-point process is realized with Weyl nodes splitting in the $k_y$ direction for $t>0$ and in the $k_x$ direction for $t<0$.

\subsubsection{With \texorpdfstring{$C^y_2\mathcal{T}$ and $C_{4z}$}{C2yT and C4z} symmetry}
Further increasing the symmetry to include $C_{4z}$ represented as $\sigma_z$ adds the constraints
\bean
d_x(k_y,-k_x,k_z)&=&-d_x(k_x,k_y,k_z)\nonumber\\
d_y(k_y,-k_x,k_z)&=&-d_y(k_x,k_y,k_z)\\
d_z(k_y,-k_x,k_z)&=&d_z(k_x,k_y,k_z).\nonumber
\eean
These constraints (with or without additional $C_{2y}\mathcal{T}$) are sufficient to make the double-Weyl point stable, with codimension 0.

\subsubsection{Inequivalent double Weyl node with \texorpdfstring{$C^y_2\mathcal{T}$}{C2yT} symmetry}
Finally we remark the existence of other, inequivalent double-Weyl nodes with $C_2 \mathcal{T}$ symmetry.
Besides higher-order nodes of the form \eqref{eqn:higher_order_dw}, the quadratic Weyl-node given by
\begin{equation}
\label{eqn:rotated_double_weyl}
    \vb{d}_2\left(\vb{k}\right)= 
    \begin{pmatrix}
        k_x^2-k_z^2\\
        k_y\\
        k_x k_z\\
    \end{pmatrix}
\end{equation}
is also inequivalent to \eqref{eqn:double_Weyl_nf}, in the sense that it cannot be reached by a K-equivalence transformation preserving the symmetry.
Even though this double-Weyl node is related to \eqref{eqn:double_Weyl_nf} by the simple transformation $d_z \leftrightarrow d_y$ and $k_z \leftrightarrow k_y$, it is qualitatively different, because both quadratic directions lie in the $k_y = 0$ symmetry-plane.
As a result, it has codimension 4 with $C_{2y}\mathcal{T}$ symmetry, with all perturbations allowed in the transverse space spanned by
\begin{equation}
    \left\langle \begin{pmatrix}
1\\
0\\
0
\end{pmatrix}, \begin{pmatrix}
0\\
0\\
1
\end{pmatrix},
\begin{pmatrix}
k_x\\
0\\
0
\end{pmatrix},
\begin{pmatrix}
0\\
0\\
k_x
\end{pmatrix}\right\rangle.
\end{equation}

\subsection{Cubic Weyl point (3-point process, co-dimension 1)}
The three merging Weyl points with total charge 1 has the normal form without symmetry
\begin{equation}
\vb{d}(\vb{k}, \vb t)=\begin{pmatrix}
k_x\\
k_y^3\\
k_z
\end{pmatrix}+
t_1\begin{pmatrix}
0\\
1\\
0
\end{pmatrix}+
t_2\begin{pmatrix}
0\\
k_y\\
0
\end{pmatrix},
\end{equation}
and in the presence of $C_{2y}\mathcal{T}$ symmetry
\begin{equation}
\vb{d}(\vb{k}, t)=\begin{pmatrix}
k_x\\
k_y^3\\
k_z
\end{pmatrix}+
t\begin{pmatrix}
0\\
k_y\\
0
\end{pmatrix},
\end{equation}
therefore, the original co-dimension 2 is reduced by the symmetry to 1.

\subsection{Quadratic Weyl point (co-dimension 1)}

A Weyl point pair creation/annihilation process has the normal form
\bean
\vb{d}(\vb{k}, t)=\begin{pmatrix}
k_x^2\\
k_y\\
k_z
\end{pmatrix}+
t\begin{pmatrix}
1\\
0\\
0
\end{pmatrix},
\eean
which does not change with the $C_{2y}\mathcal{T}$ symmetry for the two merging Weyl points on the symmetry plane. 
Therefore, the creation/annihilation process has co-dimension 1 both with and without $C_{2y}\mathcal{T}$ symmetry.

Although, the $C_{2y}\mathcal{T}$ symmetry does not change the codimension of a creation/annihilation process, the movement of the Weyl points is restricted to the symmetry plane. The colliding points can not form a non-generic degeneracy point with a quadratic dispersion in the $y$ direction.

\section{Location of the Weyl points}
We now study how the perturbation in Eq.~\eqref{eqn:double_Weyl_sym_unfolding} splits the original double Weyl point.
The Weyl points are in the \emph{configurational space} $\vb k$. The condition for the Weyl points is $\vb d_v(\vb k)=\vb 0$ and their charge is the sign of
\bean\label{eq:Jacobian}
\det(J_t(\vb k))=
\det
\begin{pmatrix}
2k_x+t_2&-2k_y&0\\
k_y&k_x&0\\
0&0&1
\end{pmatrix}=
2k_x^2+t_2k_x+2k_y^2,
\eean
where $J_t(\vb k)$ is the Jacobian of the function $\vb d(\vb k)$. From the third component, the Weyl points are always located in the $k_z=0$ plane for every control parameter.
\subsection{Out-of-plane Weyl points}
The condition of the second component $d_y=k_x k_y=0$ has two different solutions, for $k_x=0$ we get out-of-plane Weyl points. Substituting $k_x=0$ into $d_x=0$ yields
\bean\label{eq:outofplane}
-k_y^2+t_1=0.
\eean
\begin{itemize}
\item For $t_1>0$ we find a symmetric pair of Weyl points with charge sgn$\left(2k_y^2\right)=+1$.
\item At $t_1=0$ they collide at the symmetry plane.
\item For $t_1<0$ there is no Weyl point outside the symmetry plane.
\end{itemize}
\subsection{In-plane Weyl points}
The in-plane Weyl points obey the condition $k_y=0$. Substituting now $k_y=0$ into $d_x=0$ yields
\bean\label{eq:inplane}
k_x^2+t_2k_x+t_1=0.
\eean
\begin{itemize}
\item For $t_1>\frac{t_2^2}{4}$ there is no solution.
\item At $t_1=\frac{t_2^2}{4}$ we have a quadratic Weyl point with zero charge at $k_x=-\frac{t_2}{2}$.
\item For $0 < t_1<\frac{t_2^2}{4}$ there are two in-plane Weyl points with charge sgn$\left[k_x\left(2k_x+t_2\right)\right]$, so the Weyl point in the interval $-\frac{t_2}{2}<k_x<0$ has charge $-1$, the other one, corresponding to the solution with $k_x < -\frac{t_2}{2}$ has $+1$.
\end{itemize}
As $t_1$ goes through zero with $t_2\neq0$, one of the in-plane Weyl points goes through $k_x=k_y=0$, where at the same time the two out-of plane Weyl points collide. As a result, the in-plane Weyl point flips its charge. If, however, $t_1$ goes through zero with $t_2=0$, the out-of-plane Weyl points collide exactly when an in-plane Weyl point pair is created resulting in the 2-point-process.

If we want to determine the location of Weyl points with the 3-parameter perturbation in Eq.~\eqref{eq:doubleWP}, we can use the results of the 2-parameter perturbation with the shift of the configurational parameters according to Eq.~\eqref{eq:xshift} and the substitution of the connection between the control parameters described in Eq.~\eqref{eq:pullback}. The latter alone can be used to determine the control parameters corresponding to Weyl-point collisions
\begin{eqnarray}\label{eq:3Dboundary}
h_1&=&t_1-t_2t_3+t_3^2=0\\
h_1-\frac{h_2^2}{4}&=&t_1-t_2t_3+t_3^2-\frac{\left(t_2-2t_3\right)^2}{4}=t_1-\frac{t_2^2}{4}=0.\nonumber
\end{eqnarray}

\begin{figure}[h!]
	\begin{center}
		\includegraphics[width=0.5\columnwidth]{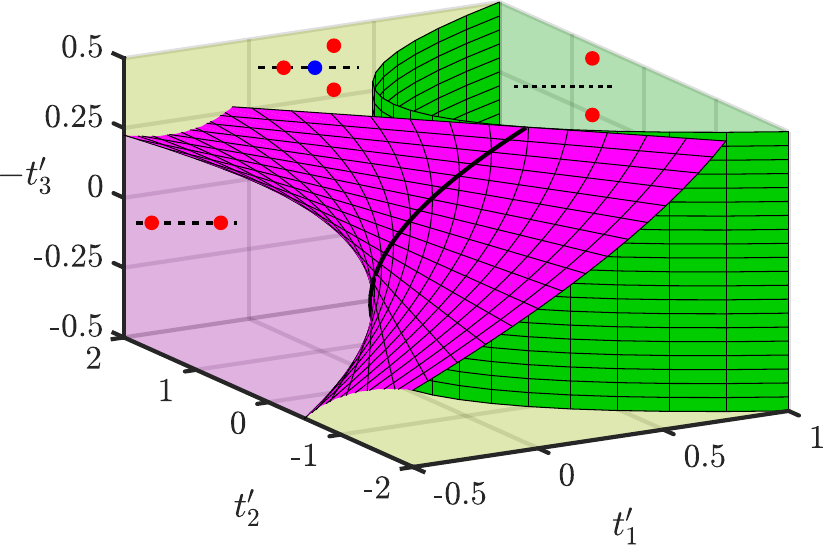}
	\end{center}
	\caption{3D Weyl phase diagram of the double Weyl point. The three-parameter perturbation doesn't add new topology compared to the two-parameter perturbation: the third parameter only increases the dimensions without making any qualitative difference of the phase diagram. The stable regions are 3 dimensional, the generic boundaries are 2 dimensional, which touch at a 1 dimensional space curve corresponding to a configuration with a double Weyl point.
	\label{fig:Weylbreaking}}
\end{figure}
\section{Weyl point configurations}
With the knowledge of the location of the Weyl points we can plot the different regions in the reduced control parameter space $(t_1,t_2)$ of Eq.~\eqref{eqn:double_Weyl_sym_unfolding} (see main text Fig.~2) and in the expanded control parameter space $(t_1,t_2,t_3)$ of Eq.~\eqref{eq:doubleWP}, see Fig.~\ref{fig:Weylbreaking}.

The stable (co-dimension 0) configurations correspond to the regions with Weyl points on the symmetry plane and/or a symmetric Weyl point pair. The two 2-Weyl point configurations are separated by the 4-Weyl point configuration, except where the phase boundaries touch (co-dimension 2) which corresponds to the 2-point process. However, the generic transition (co-dimension 1) between the two regions happens through a Weyl point pair creation on the symmetry plane and a 3-point process.

When comparing the two control parameter spaces, the expanded control parameter space introduces an additional dimension to everything: stable regions become 3D spaces, generic boundaries are 2D surfaces, and the double Weyl point exists along a 1D space curve. What remains consistent is the co-dimension of the various regions within the parameter space.

\section{Details of the \texorpdfstring{$\ce{SrSi_2}$}{SrSi2} calculations}

\subsection{Symmetry analysis of minimal model}
In \ce{SrSi_2} six double-Weyl nodes are located on the fourfold rotational axes of the cubic crystal, related to each other by the chiral cubic group ($O$).
Considering a low energy  Hamiltonian around a node at $\vb{k}_0 = (0, 0, k_{z0})$ results in the form $\vb{d}(\vb{k} - \vb{k}_0, \vb{0})$ of Eq. \eqref{eq:doubleWP}, which also has $C_{4z}$ symmetry.

The $C_{4z}$ symmetry constraint takes the form
\begin{equation}
\label{eq:C4z}
U_{4z}H(\vb{k})U_{4z}^{-1} = H(R_{z, \frac{\pi}{2}}\cdot\vb{k}),
\end{equation}
where $R_{z, \frac{\pi}{2}}$ is the real-space rotation matrix, and $U_{4z}= i \sigma_z$ acts on the internal degrees of freedom.
The $C_{4z}$ rotation operator in the basis of angular momentum $z$ eigenstates has the diagonal form $\exp\left[i (\pi/2) j_z\right]$.
Comparing with \eqref{eq:C4z} (and ignoring the overall complex phase) shows that it needs to act on states with angular momentum difference of 2, for example $j_z = 3/2,\; -1/2$. 

 In a similar way we can write down $C_{2y}\mathcal{T}=R_{y, \pi}U_{2y}U_{\mathcal{T}}\mathcal{K}$, where $R_{y, \pi}$ is the real space rotation matrix, $\mathcal{K}$ is the complex conjugation operator and $U_{2y}U_{\mathcal{T}}= -\mathbbm{1}$.
 The symmetry acts on the Hamiltonian as:
 \begin{equation}
 \label{eq:C2yT}
 (U_{2y}U_{\mathcal{T}})H^*(\textbf{k})(U_{2y}U_{\mathcal{T}})^{-1} = H(-R_{y, \pi}\cdot\vb{k}),\\
 \end{equation}
which shows that $C_{2y}\mathcal{T}$ symmetry is effectively spinless and squares to +1.

Based on the constraints on the $j_z$ eigenvalues, we choose states with a total angular momentum of $J = 3/2$ to build the $\vb{k}\cdot\vb{p}$ minimal model for \ce{SrSi_2} shown in Eq.~(2) of the main text using \texttt{qsymm}~\cite{Varjas_2018}.
The rotations are represented in the usual way for $J = 3/2$ as
\begin{equation}
    U_{\vb{n}} = \exp\left(-i \vb{n}\cdot\vb{J}\right),
\end{equation}
where $\vb{J}$ is the vector of angular momentum operators, and $\vb{n}$ is the rotation vector with $|\vb{n}|$ the rotation angle in radians and $\hat{\vb{n}}$ the rotation axis.
Time reversal has the unitary part $U_{\mathcal{T}} = \exp\left(-i \pi J_y\right)$.

\subsection{Tight binding model}

From the $\vb{k\cdot p}$ Hamiltonian (Eq.~(2) of the main text):
\begin{equation}
\label{eq:qsymmkdotp}
H(\vb{k})=\left(c_0 + c_4 \vb{k}^2\right) \mathbbm{1} + c_1 \vb{k}\cdot\vb{J} + c_2 \vb{k}\cdot\vb{S} + c_3 \sum_{i\neq j}\left(k_i k_j J_i J_j\right) + c_5 \sum_i k_i^2 J_i^2,
\end{equation}
we also obtain a tight binding model on the cubic lattice with the same key properties.
The Bloch Hamiltonian is obtained by the replacement of terms $k_i \to \sin k_i$ and $k_i^2 \to 2 (1 - \cos(k_i))$.
The resulting tight-binding model has four orbitals per site and nearest neighbour as well as next nearest neighbour hoppings.
Setting the lattice constant to 1, the Bloch Hamiltonian reads:
\begin{equation}
\label{eq:SrSitightbinding}
H(\vb{k})=c_0 \mathbbm{1} +\sum_i 2\left(1-\cos (k_i)\right)\left(c_4\mathbbm{1}+c_5J_i^2\right) + \sum_i \sin (k_i) \left(c_1J_i+c_2S_i\right) + c_3 \sum_{i\neq j}\sin (k_i) \sin (k_j) J_i J_j.
\end{equation}
The double-Weyl nodes are located at $(0, 0, \pm k_{z0})$ and symmetry related momenta, where $k_{z0}$ is the nonzero solution of
\begin{equation}
    \pm c_1 \sin k_{z0} + 2 c_5 (1 - \cos k_{0z}) = 0.
\end{equation}
For the specific choice of $c_1 = 1$, $c_5 = -1$ that we use in our calculations $k_{z0} \approx 0.9273$.

\subsection{Quasi-degenerate perturbation theory}

In order to show that the low-energy physics of the 4-band model indeed reproduces the general low-energy Hamiltonian that we constructed based only on symmetry considerations, we employ quasi-degenerate perturbation theory, restricting to the subspace of the two degenerate bands of the unperturbed Hamiltonian~\eqref{eq:qsymmkdotp}.
For simplicity we fix the parameters in the unperturbed Hamiltonian to $c_0=0$ and $c_1=1$, and treat $s$ and $B_x$ that appear in the external strain and magnetic field term
\begin{equation}
\label{eq:kdotppert}
H_{\text{pert}}(\vb{k}) = B_x J_x + s J_x^2,
\end{equation}
as small parameters.
For the above choice of parameters, the double Weyl node appears at $\vb{k}_0 = (0, 0, 1/c_5)$ as a degeneracy of the $J_z = 3/2$ and $-1/2$ bands.
We also treat the momentum components (measured from $\vb{k}_0$) as small parameters, and perform second order quasi-degenerate perturbation theory in the subspace of the two degenerate bands using the software \verb!pymablock!.~\cite{Pymablock}
The resulting effective Hamiltonian is given by
\begin{equation}
\label{eq:SrSi_pertH}
\vb{d}(\vb{k}, s, B_x)=\begin{pmatrix}
\alpha (k_{x}^{2} - k_y^2) +
\beta B_{x} k_{x} +
\frac{\sqrt{3}}{2} \left(s + \epsilon B_x^2\right)\\

\gamma k_{x} k_{y} + 
\delta B_{x} k_{y} \\

- k_{z} + 
\rho B_{x} k_{x} +
\frac{1}{2} s
\end{pmatrix},
\end{equation}
where
\begin{eqnarray}
    \alpha &=& \frac{3 \sqrt{3} \left(- c_{2} c_{3} + c_{3} - c_{5}\right)}{\left(c_2 + 1\right)\left(c_2 - 3\right)},\nonumber\\
    \beta &=& \frac{\sqrt{3} \left(- c_{2}^{2} c_{5} - 2 c_{2} c_{3} + 3 c_{2} c_{5} + 6 c_{3} - 6 c_{5}\right)}{2 \left(c_2 + 1\right)\left(c_2 - 3\right)},\nonumber\\
    \gamma &=& \frac{3 \sqrt{3} \left(c_{2}^{2} c_{3} - 2 c_{2} c_{3} + c_{2} c_{5} + c_{3} - c_{5}\right)}{\left(c_2 + 1\right)\left(c_2 - 3\right)},\\
    \delta &=& \frac{\sqrt{3} \left(- c_{2} c_{3} + 2 c_{2} c_{5} + 3 c_{3} - 3 c_{5}\right)}{\left(c_2 + 1\right)\left(c_2 - 3\right)},\nonumber\\
    \rho &=& \frac{\left(c_{2}^{2} c_{5} - 6 c_{2} c_{3} - 5 c_{2} c_{5} + 6 c_{3}\right)}{2 \left(c_2 + 1\right)\left(c_2 - 3\right)},\nonumber\\
    \epsilon &=& \frac{c_5}{c_2 + 1},\nonumber
\end{eqnarray}
and we only kept the leading order terms in the momenta and perturbations.
It can be brought to the normal form by the rescalings
\begin{eqnarray}
d'_x &=& \frac{d_x}{\alpha},\nonumber\\
d'_y &=& \frac{d_y}{\gamma},\\
d'_z &=& d_z,\nonumber
\end{eqnarray}
coordinate transformations
\begin{eqnarray}
k'_x &=& k_{x} + \frac{\delta}{\gamma} B_{x},\nonumber\\
k'_y &=& k_y,\\
k'_z &=& - k_{z} + \rho B_{x} k_{x} + \frac{s}{2},\nonumber
\end{eqnarray}
and base-change for the tuning parameters

\begin{eqnarray}
\label{eq:t1t2}
t_2 &=& B_x \left(\frac{\beta}{\alpha} - \frac{2 \delta}{\gamma}\right),\\
t_1 &=& \frac{\sqrt{3} (s + \epsilon B_x^2)}{2 \alpha} - \frac{\delta B_x^2}{\gamma} \left(\frac{\beta}{\alpha} - \frac{\delta}{\gamma}\right).\nonumber
\end{eqnarray}
Even though we neglected higher-order terms in the perturbative expansion, they can be cancelled order-by-order by applying \eqref{eqn:Ktrf} to find a suitable contact-equivalence transformation.

\subsection{Phase diagram}

\begin{figure}[h]
	\begin{center}
		\includegraphics[width=8.6cm]{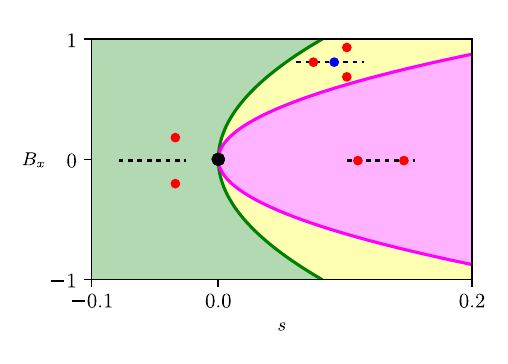}
	\end{center}
	\caption{Weyl node configurations for \ce{SrSi_2} under varying uniaxial strain $s$ and magnetic field $B_x$ in the Eq.~\eqref{eq:SrSi_pertH} model for \ce{SrSi_2}.
    Model parameters are $c_0=0$, $c_2=-0.6$, $c_3=0.75$, $c_4=1$, $c_5=-1$ (in units of $c_1$). $s$ and $B_x$ are in units of $c_1$. 
    The contours $s=P_1B_x^2$ and $s=P_2B_x^2$ are colored in magenta and green respectively.
    In the green and magenta regions we find  two Weyl points outside and inside the symmetry plane  in the yellow region we have four Weyl points. The different Weyl points configurations are shown with red/blue points. The dashed line represents the $k_y=0$ axis and the Weyl points in red (blue) carry a charge of $1$ ($-1$).
    }
 \label{fig:phasediagram_SrSi}
\end{figure}

Using the \eqref{eq:t1t2} values for the tuning parameters we can rewrite the conditions $t_1=0$ and $t_1=\frac{t_2^2}{4}$ as $s=P_1B_x^2$  and $s=P_2B_x^2$  where:
\begin{eqnarray}
P_1&=&\frac{2\alpha\delta}{\sqrt{3}\gamma}\left(\frac{\beta}{\alpha} - \frac{ \delta}{\gamma}\right)-\epsilon\\
P_2&=&\frac{\alpha}{2\sqrt{3}}\left(\left(\frac{\beta}{\alpha} - 2\frac{ \delta}{\gamma}\right)^2+4\frac{\delta}{\gamma}\left(\frac{\beta}{\alpha} - \frac{ \delta}{\gamma}\right)\right)-\epsilon.\nonumber
\end{eqnarray}
These two contours delimit different Weyl configurations in the $B_x$-$s$ plane as in Fig~\ref{fig:phasediagram_SrSi}.
The phase space is divided in regions with two positive Weyl nodes in the symmetry plane $k_y=0$ (magenta), two positive Weyl nodes outside the symmetry plane (green), and two positive nodes outside the plane and a pair of Weyl points with opposite charge inside the plane (yellow). When driving the system from the green to the magenta region a pair creation and a 3-point process are always observed, except for the fine tuned case $B_x=0$. 

\section{Phase diagram for bilayer graphene}
To obtain a phase diagram for bilayer graphene we start with the linearized Hamiltonian around the $\vb{K}$ point in the presence of skew hoppings and strain (Eq.~(3) of the main text) \cite{McCann_2013, Varlet2015}:
\begin{equation}
\label{eq:graphenelowenergy}
H(\vb{k})=\begin{pmatrix}

0           & v\pi^\dag & 0       & v_3\pi +w\\
v\pi        & 0         &\gamma_1 & 0 \\
0           &\gamma_1   & 0       & v\pi^\dag \\
v_3\pi^\dag +w& 0         &v\pi     & 0

\end{pmatrix},
\end{equation}
where $\pi=k_x+ik_y$,  $v$ is the intralayer band velocity, $\gamma_1$ is the vertical interlayer hopping, $v_3$ is the skew velocity and $w$ is the strain parameter. The Hamiltonian $H(\vb{k})$ has chiral symmetry
\begin{equation}
PH(\vb{k})P^{-1}=-H(\vb{k}),
\end{equation}
with
\begin{equation}
    P=\begin{pmatrix}
    1&0&0&0\\
    0&-1&0&0\\
    0&0&1&0\\
    0&0&0&-1
    \end{pmatrix}.
\end{equation}
The chiral symmetry protects the degeneracy points of bilayer graphene and ensures the conservation of the total winding number.

Furthermore, the symmetry forces the degeneracy between the second and third band to be at zero energy.  This means that  requiring that the Hamiltonian has zero energy eigenstates, i.e., $H(k_x, k_y)\ket{\Psi}=0$ with the general $\ket{\Psi}= (\psi_1,\psi_2,\psi_3,\psi_4)^T$, leads to a system of equation for the existence of degeneracy points :
\begin{eqnarray}
      &v\pi^\ast\psi_2+(v_3\pi+w)\psi_4=0,\nonumber\\
      &v\pi\psi_1+\gamma_1\psi_3=0,\\
      &\gamma_1\psi_2+v\pi^\ast\psi_4=0,\nonumber\\
      &(v_3\pi^\ast+w)\psi_1+v\pi\psi_3=0,\nonumber
\end{eqnarray}
where $\pi^*=k_x-ik_y$ denotes the complex conjugate of $\pi$. The second equation gives  $\psi_3=-\frac{v\pi}{\gamma_1}\psi_1$, and substituting this to the fourth equation yields
\begin{equation}
    (v_3\pi^\ast+w)\psi_1-\frac{v^2\pi^2}{\gamma_1}\psi_1=0.
\end{equation}
which gives
\begin{equation}\label{eq:zeroenergy}
-\frac{v^2}{\gamma_1}\pi^2+v_3\pi^\ast+w=0
\end{equation}
for $\psi_1\neq 0$. Similarly, the first and third equation gives the conjugate of Eq.~\eqref{eq:zeroenergy}. The real-imaginary decomposition yields
\begin{eqnarray}
\label{eq:grapheneDiracequationsa}
&-\frac{v^2}{\gamma_1}\left(k_x^2-k_y^2-\frac{\gamma_1 w}{v^2}-\frac{\gamma_1v_3}{v^2}k_x\right)=0,\\
\label{eq:grapheneDiracequationsb}
&-\frac{2v^2}{\gamma_1}\left(k_xk_y+\frac{\gamma_1v_3}{2v^2}k_y\right)=0.\nonumber
\end{eqnarray}
This set of equations is equivalent to the equation $\vb{d}_0(\vb{k},\vb{t})=0$ up to constant factors, where $\vb{d}_0$ is defined in Eq.~\eqref{eq:doubleWP} and $\vb{k}$ is the configurational parameter. The control parameters read
\begin{eqnarray}\label{eq:bilayercontrol}
t_1&=&-\frac{\gamma_1w}{v^2},\nonumber\\
t_2&=&-\frac{\gamma_1v_3}{v^2},\\
t_3&=&\frac{\gamma_1v_3}{2v^2}.\nonumber
\end{eqnarray}

Eq.~\eqref{eq:grapheneDiracequationsa} and Eq.~\eqref{eq:grapheneDiracequationsb} give the number and the position of the Dirac points around $\vb{K}$ for every value of the skew velocity $v_3$ and of the strain parameter $w$.

To draw the phase diagram in the $v_3$-$w$ plane we substitute Eq.~\eqref{eq:bilayercontrol} to the equations of the phase boundaries defined in Eq.~\eqref{eq:3Dboundary}:
\begin{eqnarray}
w &=& \frac{3}{4}\frac{\gamma_1 v_3^2}{v^2}\\
w &=& -\frac{1}{4}\frac{\gamma_1 v_3^2}{v^2}.
\end{eqnarray}
 \begin{figure}[h]
	\begin{center}
		\includegraphics[width=8.6cm]{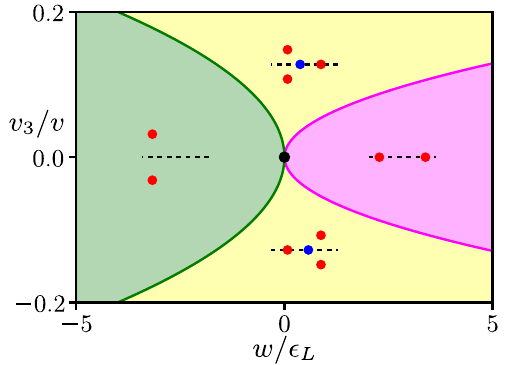}
	\end{center}
	\caption{Dirac point configurations near the $K$ point of bilayer graphene \eqref{eq:graphenelowenergy} in the presence of skew velocity $v_3$ and strain $w$. $w$ is measured in units of the Lifshitz energy $\epsilon_L= 1$ meV and $v_3$ is measured in units of the intralayer velocity $v=10^6 m/s$. 
 The green and magenta lines are respectively the analytic contours $w=-\frac{1}{4}\frac{\gamma_1 v_3^2}{v^2}$ and $w=\frac{3}{4}\frac{\gamma_1 v_3^2}{v^2}$, for $\gamma_1=0.4$ eV, which delimit the area of where four Dirac points are present (yellow). 
The different Dirac points configurations are shown with red/blue points. The dashed line represents the $p_y=0$ axis and the Dirac points in red (blue) carry a winding number of $1$ ($-1$).}
   \label{fig:phase_graphene}
 \end{figure}
These two contours delimit three regions in the phase diagram of bilayer graphene in Fig. \ref{fig:phase_graphene}.
There is one region (green) with two out of symmetry axis ($k_y=0$) Dirac points and one (magenta) with  two in axis cones.  Each of these cones carry a charge of $+1$.
In between these two regions we have four Dirac cones: two points inside the symmetry axis with opposite charge $\pm 1$ and one pair outside the axis with the same charge $+1$.

\bibliography{references}